\documentclass[twocolumn,amsmath,amssymb]{revtex4-1}

\usepackage[dvips]{graphicx}
\usepackage[utf8]{inputenc}
\usepackage[T1]{fontenc}
\usepackage{mathptmx}
\usepackage{etoolbox}
\usepackage{siunitx}
\usepackage{color}
\usepackage{upgreek}
\usepackage{xspace}

\newenvironment{fequation*}{\empheq[box=\fbox]{equation*}}{\endempheq}
\newenvironment{falign*}{\empheq[box=\fbox]{align*}}{\endempheq}
\newenvironment{fmultline*}{\empheq[box=\fbox]{multline*}}{\endempheq}
\renewcommand{\bm}[1]{\boldsymbol{\mathbf{#1}}}
\newcommand{\ud}{\mathrm{d}}
\newcommand{\bra}{\left\langle}
\newcommand{\ket}{\right\rangle}

\renewcommand{\Re}{\operatorname{Re}}
\renewcommand{\Im}{\operatorname{Im}}
\DeclareMathOperator{\sinc}{sinc}

\newcommand{\ie}{i.e.\@\xspace}

\usepackage[normalem]{ulem}

\begin{document}

   \title{Characterization of ejecta in shock experiments with multiple light scattering}

   \author{J.A. Don Jayamanne}
   \affiliation{CEA DIF, Bruyères-le-Châtel, 91297 Arpajon Cedex, France}
   \affiliation{Institut Langevin, ESPCI Paris, PSL University, CNRS, 75005 Paris, France}
   \author{J.-R. Burie}
   \email{jean-rene.burie@cea.fr}
   \author{O. Durand}
   \affiliation{CEA DIF, Bruyères-le-Châtel, 91297 Arpajon Cedex, France}

   \author{R. Pierrat}
   \affiliation{Institut Langevin, ESPCI Paris, PSL University, CNRS, 75005 Paris, France}
   \author{R. Carminati}
   \email{remi.carminati@espci.psl.eu}
   \affiliation{Institut Langevin, ESPCI Paris, PSL University, CNRS, 75005 Paris, France}
   \affiliation{Institut d'Optique Graduate School, Paris-Saclay University, 91127 Palaiseau, France}

   \date{\today}

   \begin{abstract}
      Upon impact, the free surface of a solid metal may eject a cloud of fast and fine particles. Photon Doppler 
      Velocimetry (PDV) is one of the optical diagnostics used to characterize these ejecta. Although the technique 
      provides a direct way to estimate the particle velocities in the single scattering regime, it has been shown
      that multiple scattering cannot be neglected in real ejecta. Here we derive a model for PDV measurements 
      starting from first principles of wave scattering. We establish rigorously the relationship between the specific 
      intensity and the measured signal, as well as the radiative transport equation (RTE) that describes the evolution 
      of the specific intensity upon scattering and absorption in a dynamic ejecta, including the effects of inelastic
      scattering and inhomogenities in the optical properties. We also establish 
      rigorously the connection between the Monte-Carlo scheme used for numerical simulations and the solution to the RTE. 
      Using numerical simulations, we demonstrate the crucial role of multiple scattering and inhomogeneities in the particle 
      density and size-velocity distribution. These results could substantially impact the analysis of ejecta by PDV.
   \end{abstract}

   \maketitle

   \section{Introduction}\label{introduction}
   % ====================

   The extreme conditions of shock-compression experiments make it possible to observe unique physical phenomena. Although 
   they have specific applications, for example in inertial fusion\cite{dimonte_ejecta_2013}, from a purely physical point of 
   view, these regimes are worth investigating in their own right. Under the correct impact conditions, shockwaves inside a 
   material cause it to partially melt and its free surface to eject a cloud of particles called ejecta. It has been shown 
   that ejecta creation is mainly due to surface imperfections of machined materials~\cite{asay_ejection_1976,
   andriot_ejection_1982}. Since then, ejecta formation has been intensively studied~\cite{buttler_foreword_2017} with the 
   goal of predicting both the total amount of ejected mass and the size-velocity distribution of the ejecta. This effort was 
   twofold.
      
   On the ejecta modelling side, there have been advances in theory~\cite{buttler_unstable_2012,dimonte_ejecta_2013}
   showing that ejecta are a limiting case of Richtmyer-Meshkov instabilities. Numerically, continuum simulations
   ~\cite{fung_ejecta_2013} and more recently molecular dynamics simulations ~\cite{durand_large-scale_2012,durand_power_2013,
   durand_mass-velocity_2015} have made it possible to determine expected size-velocity distributions. Parallel to these 
   advances, a large number of diagnostics were developed and used in experiments to obtain as much information 
   as possible on the ejecta~\cite{buttler_foreword_2017,buttler_understanding_2021}. Among other optical diagnostics, Photon 
   Doppler Velocimetry (PDV) has played a key role thanks to its ability to simultaneously measure the velocities of several 
   targets, and to the trade-off between a fine temporal resolution and a broad range of accessible velocities
   ~\cite{STRAND-2006,mercier_photonic_2006}. However, current analysis of PDV signals relies on the single scattering hypothesis (light collected 
   is assumed to be scattered only once in the medium), which provides a direct relationship between the measured Doppler 
   shifts and the velocity of the scatterers. Since in practice the thickness of the ejecta can 
   largely exceed the photon scattering mean-free path, accounting for multiple scattering is of utmost importance for a 
   quantitative analysis of the 
   experiments.

   In the last decade, several models have been developed to account for multiple scattering
   ~\cite{andriyash_optoheterodyne_2016,franzkowiak_multiple_2018,andriyash_application_2018,andriyash_photon_2020,
   kondratev_application_2020,shi_reconstruction_2021,andriyash_simultaneous_2022,shi_reconstruction_2022}.
   A key point is to assume that the PDV spectrograms correspond to the specific intensity detected by the probe, and that the
   specific intensity obeys the Radiative Transfer Equation (RTE)~\cite{CHANDRASEKHAR-1950}. An advantage of the RTE is its
   ability to deal with complex particle size and velocity distributions, while being solvable numerically in realistic 
   geometries. The use of the RTE raises several basic issues that remains to be clarified. Although a phenomenological 
   approach is widely used to introduce the specific intensity and the RTE, a rigorous framework based on the wave equation
   and a statistical description of the scattering medium is available~\cite{APRESYAN-1996,carminati_principles_2021}. This 
   theoretical framework should allow one to establish rigorously the relationship between the specific intensity solution to 
   the RTE and the spectrogram deduced from the PDV signal. Moreover, the RTE that is used is usually modified to account for 
   Doppler shifts and inhomogeneities in the scattering properties of the medium. This generalized form of the RTE also 
   deserves to be derived in more than an intuitive way. Finally, Monte-Carlo simulations are widely used
   ~\cite{franzkowiak_multiple_2018,shi_reconstruction_2021,shi_reconstruction_2022}, but the fact that they provide 
   solutions to the (generalized) RTE is often overlooked. Proving that the Monte-Carlo approach amounts to solving the RTE 
   would fill a gap in the chain connecting simulations to experiments.

   The purpose of this work is to answer these questions, in order to provide a complete theory supporting the analysis of 
   PDV experiments in ejecta. To proceed, we first establish a rigorous connection between the PDV signal and the specific 
   intensity scattered by the ejecta and collected by the probe. Then, we derive the RTE from the wave equation, taking into 
   account all important features of shock ejecta such as the size and velocity distributions of the scatterers, and 
   heterogeneities in the number density. Finally, we establish a random walk scheme that naturally supports the connection 
   between Monte-Carlo simulations and the solution to the RTE. The model is used to demonstrate unambiguously the importance 
   of multiple scattering in shock ejecta.

   The paper is organized as follows. In Sec.~\ref{pdv_multiple_scattering}, we describe a typical shock experiment
   including the PDV setup used to characterize the ejecta. We also explain how a PDV spectrogram can estimate accurately
   the particle velocity distribution under the assumption of single scattering of the probe light.
   Section~\ref{spectrogram_specific_intens_link} is dedicated to the analysis of the multiple scattering regime, and
   the link between the spectrograms computed from PDV measurements and the specific intensity of the collected light.
   The derivation of the RTE describing multiple scattering of light scattering in ejecta is reported in
   Sec.~\ref{rte}. Finally, the justification of the Monte-Carlo model is presented in Sec.~\ref{numerics},
   together with some applications to the analysis of real spectrograms.

   \section{Shock experiments and photon doppler velocimetry in the single scattering regime}\label{pdv_multiple_scattering}
   % ========================================================================================

\begin{figure}[!htb]
      \centering
      \includegraphics[width=0.8\linewidth]{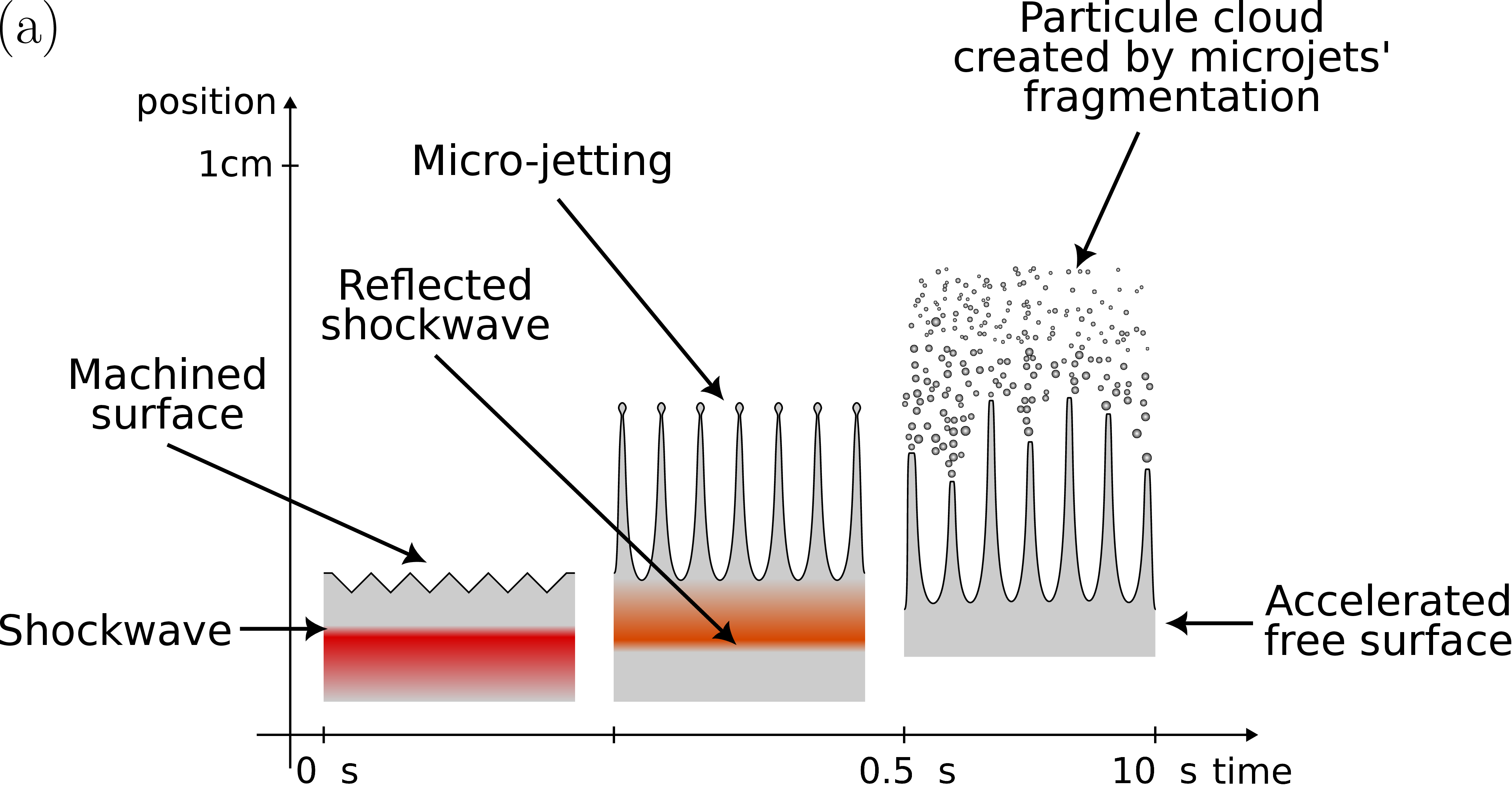}
      \includegraphics[width=0.8\linewidth]{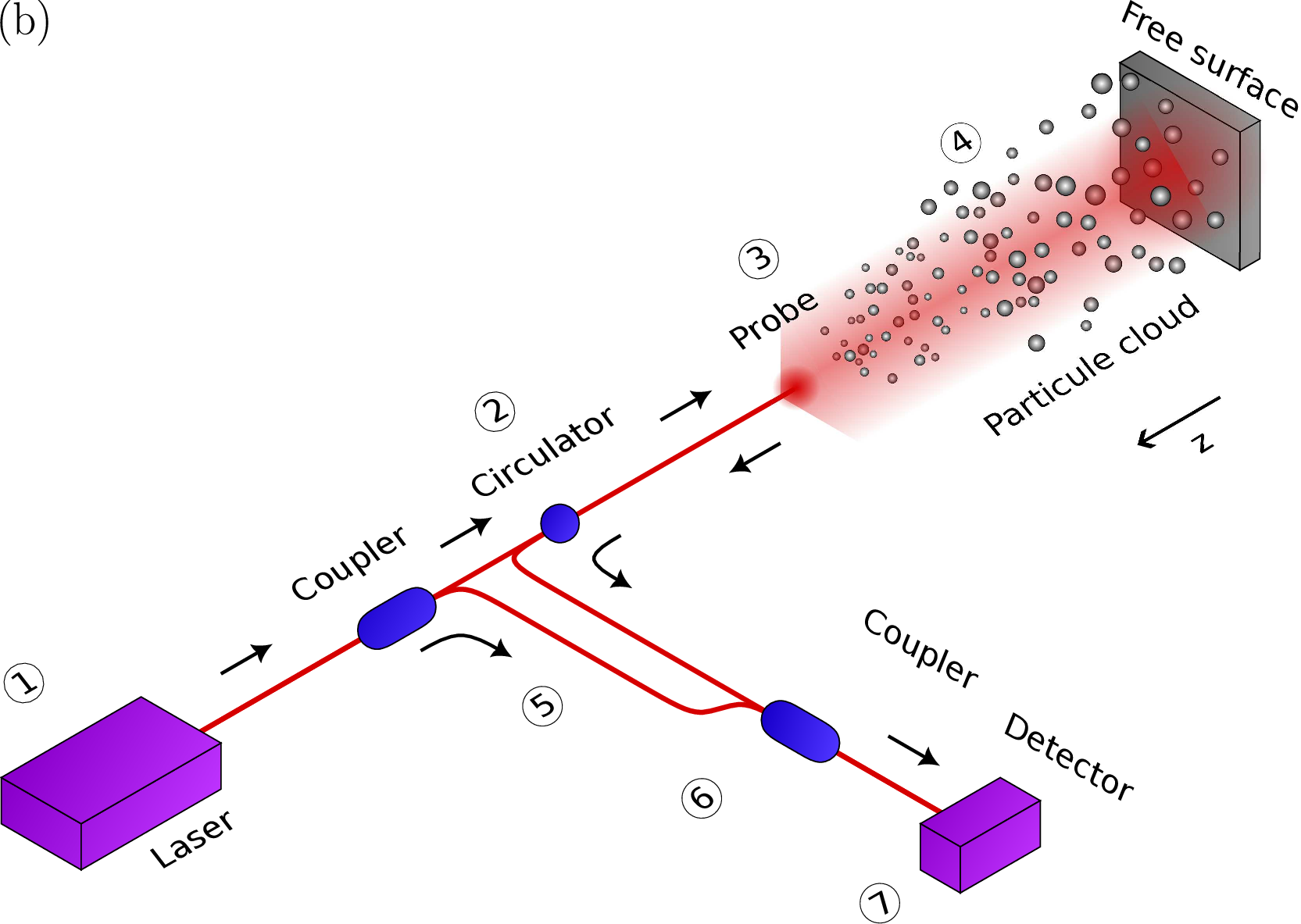}
      \includegraphics[width=0.8\linewidth]{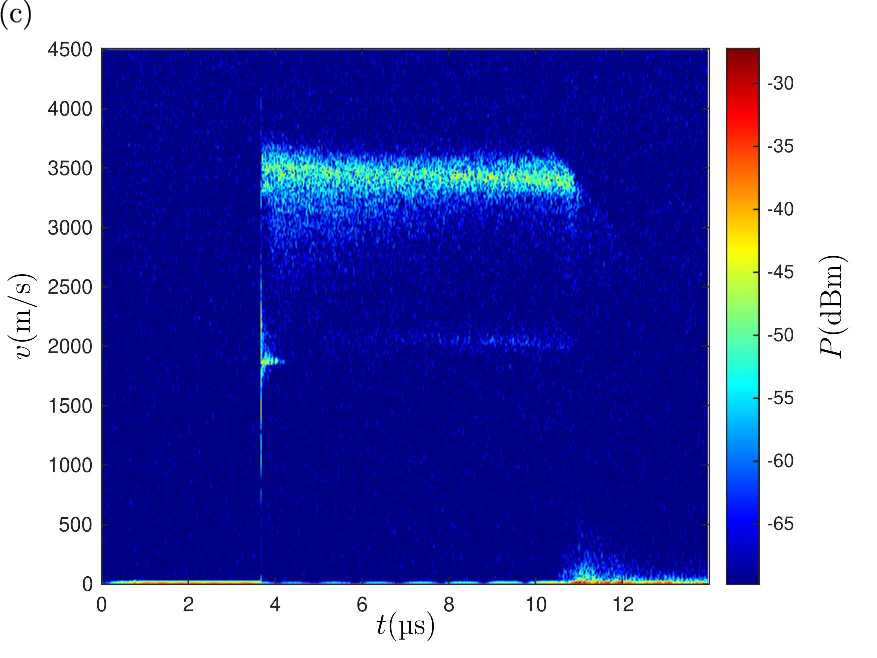}
      \caption{(a) Illustration of the micro-jet mechanism in a typical shock ejecta experiment. Upon reaching the
      machined free surface, the shock wave first comes into contact with the inwardly directed grooves. Under right
      angle conditions, the shock wave is reflected and the inward grooves become outward micro-jets. Due to the velocity
      gap between the jet-heads and the free surface, the micro-jets are stretched until surface tension is no longer
      sufficient to hold matter together and fragmentation begins. This results in the creation of an ejecta. (b)
      Schematic representation of a typical shock-loaded experiment with a PDV setup. The probe illuminates the
      ejecta and the free surface with a highly collimated laser beam (numerical aperture of \SI{4.2}{\milli rad} and
      pupil size $\phi_p=\SI{1.3}{\micro m}$). The backscattered field is collected by the probe as the
      measuring arm and interferes with the reference arm at the detector. The beating signal is registered with a high
      bandwidth oscilloscope before being analyzed. (c) Spectrogram of a tin micro-jetting experiment under pyrotechnic
      shock at $P=\SI{25}{\giga Pa}$. The free surface was engraved with $\SI{25}{\micro m}\times\SI{8}{\micro m}$
      grooves. Independent Asay window measures gave an estimated surface mass $M_s = \SI{5}{\milli g / \centi m^2}$.}
      \label{setup_spectro}
   \end{figure}

   In this section, we recall some standard results for the characterization of shock experiments using PDV in the
   single scattering regime. In a typical shock experiment, a shockwave is released by impact in a material. Under this
   extreme excitation, the material partially melts and the shockwave interacting with the irregularities of the free surface 
   generates micro-jets of matter. These micro-jets extend before fragmenting and forming a cloud of fast and fine particles, 
   referred to as the ejecta [see Fig.~\ref{setup_spectro}\,(a)].

   Ejecta are usually characterized optically using interferometric techniques such as a
   PDV~\cite{STRAND-2006,mercier_photonic_2006}. The usual setup is a Michelson interferometer, in which the reference beam interferes 
   with the scattered light coming from a measuring arm, as represented schematically in Fig.~\ref{setup_spectro}\,(b).
   Since the particles and the free surface are moving, the scattered field is slightly Doppler shifted in frequency compared
   to the light in the reference arm.This creates a beating signal on a photodiode detector, that allows one to
   directly measure the Doppler spectrum of the scattered field. In practice, a time-frequency analysis of the signal is 
   performed, resulting in a spectrogram as that shown in Fig.~\ref{setup_spectro}\,(c). The usual representation directly 
   transforms frequency shifts into particle velocities, under the fundamental assumption of single scattering. As a result, 
   a vertical line of the spectrogram is understood as the velocity distribution of the particles at a given time. 
  
   Let us briefly summarize the main steps leading to a theoretical description of the spectrogram in the single scattering 
   regime. The measured signal corresponds to the intensity $I(t)=|\bar{E}(t)|^2$, where $\bar{E}(t)$ is the analytical signal 
   associated to the real-valued field $E(t)$ describing the light received on the detector. Focusing on the time dependence,
   the analytic field is defined as
   \begin{equation}\label{analytic_field}
      \bar{E}(t)=\int_{0}^{+\infty} E(\omega)\exp(-i\omega t)\frac{\ud\omega}{\pi}
   \end{equation}
   where $E(\omega)$ is the Fourier transform of $E(t)$. We use a scalar description of the light field, leaving aside
   polarization effects. Considering that the illuminating field is monochromatic at frequency $\omega_0$, the associated
   analytical signal is $\bar{E}_0(t)=\mathcal{A}_0\exp(-i\omega_0t)$. The analytical signal associated to the scattered
   field can be written
   \begin{equation}\label{single_scattered_field}
      \bar{E}_s(t)=\sum_{j=1}^{N(t)}\mathcal{A}_j(t)\exp\left\{-i\left[\omega_0+\delta\omega_j(t)\right]t\right\}
   \end{equation}
   where $N(t)$ is the number of particles, $\mathcal{A}_j(t)$ is the amplitude of the field scattered by particle number $j$,
   and $\delta\omega_j(t)$ is the corresponding Doppler shift. These quantities evolve on a time scale $T_c$, which 
   corresponds to the characteristic time of changes in the statistical properties of the cloud of particles. The field also 
   contains the time scale $\delta T = 2\pi/\delta\omega$, where $\delta\omega$ is the typical Doppler shift. We note that 
   these time scales are expected to be much larger than the period $T_0=2\pi/\omega_0$ of the incident light field.
   
   The intensity received by the photodiode is
   \begin{align}
      I(t)=\left|\bar{E}_0(t)+\bar{E}_s(t)\right|^2
      \simeq \left|\mathcal{A}_0\right|^2+2\Re\left[\bar{E}_s(t)\bar{E}_0^*(t)\right] \,
   \end{align}
   where the superscript $*$ denotes the complex conjugate. We have dropped the term $\left|\bar{E}_s(t)\right|^2$ since the 
   amplitude of the scattered field is much smaller than the amplitude of the illuminating field. The first term
   $\left|\mathcal{A}_0\right|^2$ is a DC component that can be ignored, and the second term defines the real-valued useful
   detected signal
   \begin{equation}\label{useful_detected_signal}
      {\cal I}(t)=2\Re\left[\bar{E}_s(t)\bar{E}_0^*(t)\right].
   \end{equation}
   Using Eq.~(\ref{single_scattered_field}), we immediately find that the detected signal is
   \begin{equation}\label{signal_single}
      {\cal I}(t) = 2\left|\mathcal{A}_0\right|\sum_{j=1}^{N(t)}\left|\mathcal{A}_j(t)\right|\cos\left\{\left[\delta\omega_j
      (t)\right]t\right\}.
   \end{equation}
   In practice, a Short Term Fourier Transform (STFT) is performed on the signal, defining the spectrogram as
   \begin{equation}\label{spectrogram}
     S(t,\omega)=\left|\int \mathcal{I}(\tau)w(\tau-t)\exp(i\omega\tau)\ud\tau\right|^2
   \end{equation}
   where $w(t)$ is a time window function centered at $t=0$. Various shapes~\cite{harris_use_1978} can be used for $w(t)$, 
   the key parameter being its duration $T_w$ that is chosen to clearly separate the time scales $T_c$ and $\delta T$ 
   characterizing the ejecta dynamics. To this aim, the inequalities $T_0\ll T_d\ll \delta T\ll T_w\ll T_c$ need to be 
   satisfied, where $T_d$ is the time scale associated to the digitizer bandwidth. Typical orders of magnitudes for usual 
   experiments are $T_0=\SI{e-15}{s}$, $T_d=\SI{e-10}{s}$, $\delta T=\SI{e-9}{s}$, $T_w=\SI{e-8}{s}$ and
   $T_c=\SI{e-6}{s}$. For illustration purposes, we consider here a rectangular function for $w(t)$, with unit amplitude 
   and width $T_w$. This choice is not restrictive and allows us to describe the main features of PDV spectrograms. In these 
   conditions, from Eqs.~(\ref{signal_single}) and (\ref{spectrogram}), we can show that
   \begin{widetext}
      \begin{equation}
         S(t,\omega)\simeq\left|
            \sum_{j=1}^{N(t)}\left|\mathcal{A}_0\right|\left|\mathcal{A}_j(t)\right|T_w\left(
               \sinc\left\{\frac{[\omega+\delta\omega_j(t)] T_w}{2}\right\}
               +\sinc\left\{\frac{[\omega-\delta\omega_j(t)] T_w}{2}\right\}
            \right)
         \right|^2
      \end{equation}
   \end{widetext}
   where $\sinc(x)=\sin(x)/x$. In addition, since $T_w$ is large compared to $T_0$ and $\delta T$, the $\sinc$ functions
   can be approximated by Dirac delta functions, leading to
   \begin{multline} \label{spectrogram_delta}
      S(t,\omega)\simeq\pi^2\left|\mathcal{A}_0\right|^2\sum_{j=1}^{N(t)}\left|\mathcal{A}_j(t)\right|^2
      \\\times
      \left\{\left|\delta[\omega+\delta\omega_j(t)]\right|^2+\left|\delta[\omega-\delta\omega_j(t)]\right|^2\right\} \, .
   \end{multline}
   In the single scattering regime, a mapping between the frequency shifts and the particle velocities is obtained
   by writing that for a scattering event occuring on particle $j$ with velocity $\bm{v}_j(t)$, the Doppler shift\cite{briers_laser_1996} is
   \begin{equation}\label{doppler}
      \delta\omega_j(t)=\bm{v}_j(t)\cdot(\bm{k}_s-\bm{k}_i) \, .
   \end{equation}
   In this expression $\bm{k}_s$ and $\bm{k}_i$ are the scattered and incoming wavevectors respectively, such that
   $|\bm{k}_s|=|\bm{k}_i|=k_0=\omega_0/c=2\pi/\lambda$, with $\lambda$ the incident wavelength. In the configuration
   presented in Fig.~\ref{setup_spectro}\,(b), where the backscattered light is detected in direction $\bm{k}_s=-\bm{k}_i$,
   the Doppler shift is simply $\delta\omega_j(t)= 4\pi v_j(t)/\lambda$. Inserting this expression into
   Eq.~(\ref{spectrogram_delta}),
   we immediately see that in this configuration we only have access to the absolute value $v_j(t)$ of the particle 
   velocities.

   The analysis above relies on the assumption of single scattering. In a statistically homogeneous cloud with longitudinal 
   size $L$, this assumption holds as long as the optical thickness $b=L/\ell_s\ll 1$, where $\ell_s$ is the scattering mean 
   free path. In an inhomogeneous cloud, the criterion becomes
   \begin{equation}
      b=\int \frac{\ud z}{\ell_s(z)}\ll 1
   \end{equation}
   where $z$ denotes the coordinate along the direction perpendicular to the free surface. In practice, many situations of
   interest rather correspond to $b>1$, and even $b \gg 1$~\cite{franzkowiak_pdv-based_2018,franzkowiak_multiple_2018,
   buttler_understanding_2021}. For example, we will describe in Sec.~\ref{numerics} a situation with $b=42$, 
   corresponding to a real experiment. Obviously, analyzing experiments in the deep multiple scattering regime with the 
   formalism described above is irrelevant, and leads to artefacts such as apparent velocities smaller than the free surface 
   velocity~\cite{franzkowiak_multiple_2018}, as seen in Fig.~\ref{setup_spectro}\,(c). Establishing a rigorous treatment
   of spectrograms in the multiple scattering regime is the purpose of the next sections.

   \section{Spectrogram and specific intensity}\label{spectrogram_specific_intens_link}
   % ==========================================

   The RTE appears as a natural tool to describe light scattering in ejecta in the presence of multiple scattering. Since the
   RTE is a transport equation for the specific intensity, we will start by establishing a rigorous relationship between the
   specific intensity and the spectrogram. The key step in defining the spectrogram from the measured PDV signal
   in Eq.~(\ref{spectrogram}) is a time analysis, and we will focus on the time dependence and omit the space dependence of 
   the field in the notations in the beginning of this section.
  
   For a rectangular time window function with width $T_w$ and unit amplitude, the spectrogram in Eq.~(\ref{spectrogram}) 
   becomes
   \begin{equation}\label{spectrogram_theo}
      S(t,\omega)=\left|\int_{t-T_w/2}^{t+T_w/2} {\cal I}(\tau)\exp(i\omega\tau)\ud\tau\right|^2.
   \end{equation}
   In statistical optics, the specific intensity is defined as the time and space Wigner transform of the field
   ~\cite{BARABANENKOV-1969,RYTOV-1989,APRESYAN-1996,carminati_principles_2021}. Focusing on the time dependence,
   the specific intensity of the scattered field is
   \begin{equation}\label{specific_intensity_theo}
      I_s(t,\omega)=\int \bra\bar{E}_s\left(t+\frac{\tau}{2}\right)\bar{E}_s^*\left(t-\frac{\tau}{2}\right)\ket
      \exp(i\omega \tau)\ud \tau \, ,
   \end{equation}
   where the notation $\bra\cdots\ket$ denotes a statistical average over an ensemble of realizations of the scattering 
   medium.
   
   Since we no longer assume a single scattering regime, we cannot use Eqs.~(\ref{single_scattered_field})~
   and~(\ref{signal_single}) anymore. Nonetheless, the definition given for $\mathcal{I}(t)$ in
   Eq.~(\ref{useful_detected_signal}) remains valid. To link both expressions, we start by introducing the Fourier transform
   of $\mathcal{I}(t)$ in Eq.~(\ref{spectrogram_theo}), and rewrite the spectrogram in the form
   \begin{multline}
      S(t,\omega)=\int_{-\infty}^{+\infty}\int_{-\infty}^{+\infty} \mathcal{I}\left(\omega'+\frac{\Omega}{2}\right)\mathcal{I}
      ^*\left(\omega'-\frac{\Omega}{2}\right)\exp\left(-i\Omega t\right)
   \\\times
      T_w\sinc\left(\frac{\omega-\omega'-\Omega/2}{2}T_w\right)
   \\\times
      T_w\sinc\left(\frac{\omega-\omega' +\Omega/2}{2}T_w\right)\frac{\ud \omega'}{2\pi}\frac{\ud \Omega}{2\pi}.
   \end{multline}
   The frequency $\Omega$ associated to the time variable $t$ encodes the slow variations of $\mathcal{I}$ on the scale of $T_c$.
   Since $T_w\ll T_c$ we have $\Omega T_w\ll 1$, and we can simplify the expression which becomes
   \begin{multline}\label{frequency_filtering}
      S(t,\omega)=\int_{-\infty}^{+\infty} \frac{\ud \omega'}{2\pi} \ T_w^2\sinc^2\left(\frac{\omega-\omega'}{2}T_w\right)
   \\\times
      \int_{-2\pi/T_w}^{+2\pi/T_w}\frac{\ud \Omega}{2\pi} \ \mathcal{I}\left(\omega'+\frac{\Omega}{2}\right)\mathcal{I}^*\left(\omega' -\frac{\Omega}{2}\right)\exp\left(-i\Omega t\right).
   \end{multline}
   Transforming back to the time domain for ${\cal I}(t)$, we find that
   \begin{multline}\label{time_averaging}
      S(t,\omega)=T_w\int_{-\infty}^{+\infty}\frac{\ud \omega'}{2\pi} \ \sinc^2\left(\frac{\omega-\omega'}{2}T_w\right)
   \\\times 
      \int_{t'=t-T_w/2}^{t+T_w/2} \int_{\tau=-\infty}^{\infty} \ud \tau \ud t' \ \mathcal{I}\left(t'+\frac{\tau}{2}\right)
      \mathcal{I}^*\left(t'-\frac{\tau}{2}\right)\exp(i\omega' \tau).
   \end{multline}
   The gating window having a width $T_w$ large compared to $\delta T$ and $T_0$ and small compared to $T_c$, the second
   integral can be replaced by a statistical average over the configurations of the disordered cloud by invoking ergodicity. 
   This leads to
   \begin{multline}\label{configurational_averaging}
      S(t,\omega)=T_w^2\int_{-\infty}^{+\infty}\frac{\ud \omega'}{2\pi} \ \sinc^2\left(\frac{\omega-\omega'}{2}T_w\right)
   \\\times
      \int_{-\infty}^{+\infty}  \ud\tau \ \bra \mathcal{I}\left(t+\frac{\tau}{2}\right)\mathcal{I}^*\left(t-\frac{\tau}{2}\right)\ket\exp(i\omega'\tau).
   \end{multline}
   Since $\delta T\ll T_w$, the $\sinc^2$ function can be replaced by a Dirac delta function, which leads to
   \begin{equation}\label{narrow_sinc}
      S(t,\omega)=T_w\int_{-\infty}^{+\infty}
      \bra \mathcal{I}\left(t+\frac{\tau}{2}\right)\mathcal{I}^*\left(t-\frac{\tau}{2}\right)\ket\exp(i\omega\tau)\ud\tau.
   \end{equation}
   Making use of Eq.~(\ref{useful_detected_signal}) and keeping only the terms corresponding
   to frequencies on the order of $\delta\omega$, the spectrogram becomes
   \begin{equation}
      S(t,\omega)=T_w\left|\mathcal{A}_0\right|^2\left[I_s(t,\omega_0+\omega)+I_s(t,\omega_0-\omega)\right],
   \end{equation}
   with $I_s(t,\omega)$ the specific intensity defined in Eq.~(\ref{specific_intensity_theo}). This expression connects
   the spectrogram to the specific intensity. The full expression also exhibits a dependence on space variables (that
   were not used in the above derivation). Indeed, the specific intensity $I_s(\bm{r},\bm{u},t,\omega)$ is defined by the 
   relation
   \begin{multline}\label{specific_intensity_scattered}
      \delta(k-k_R)I_s(\bm{r},\bm{u},t,\omega)
   \\
      =\int
      \bra\bar{E}_s\left(\bm{r}+\frac{\bm{\uprho}}{2},t+\frac{\tau}{2}\right)
      \bar{E}_s^*\left(\bm{r}-\frac{\bm{\uprho}}{2},t-\frac{\tau}{2}\right)\ket
   \\\times
      \exp(-ik\bm{u}\cdot\bm{\uprho}+i\omega \tau)\ud \bm{\uprho}\ud \tau
   \end{multline}
   where $k_R$ is the real part of the effective wavevector in the scattering medium and $\bm{u}$ is a unit vector. 
   The meaning of this definition, and the fact that $I_s(\bm{r},\bm{u},t,\omega)$ is the solution to the RTE, 
   will be clarified in Sec.~\ref{rte}. In the large wavelength limit, it can be shown that $I_s$ is always positive and can 
   be interpreted as the radiative flux at position $\bm{r}$, in direction $\bm{u}$, at time $t$ and at frequency 
   $\omega$~\cite{MANDEL-1995}. From this expression, the full spectrogram finally takes the form
   \begin{multline}\label{spectrogram_specific_intensity}
      \delta(k-k_R)S(t,\omega)=T_w\left|\mathcal{A}_0\right|^2
   \\\times
      \int_G\left[I_s(\bm{r},\bm{u},t,\omega_0+\omega)+I_s(\bm{r},\bm{u},t,\omega_0-\omega)\right]\bm{u}\cdot\bm{n} \, \ud
      \bm{u}\ud\bm{r}.
   \end{multline}
   Here $G$ is the etendue of detection of the photodiode (surface of the photodiode and angular aperture), $\ud\bm{u}$ means 
   integration over the solid angle, and $\bm{n}$ is the normal to the detector surface. In this expression, the specific 
   intensity for the scattered field can be replaced by the specific intensity for the full field $I(\bm{r},\bm{u},t,
   \omega_0\pm\omega)$ since the incident field ${E}_0$ does not play any role for detection in the backward direction.
   It is given by
   \begin{multline}\label{specific_intensity_full}
      \delta(k-k_R)I(\bm{r},\bm{u},t,\omega)
   \\
      =\int
      \bra\bar{E}\left(\bm{r}+\frac{\bm{\uprho}}{2},t+\frac{\tau}{2}\right)
      \bar{E}^*\left(\bm{r}-\frac{\bm{\uprho}}{2},t-\frac{\tau}{2}\right)\ket
   \\\times
      \exp(-ik\bm{u}\cdot\bm{\uprho}+i\omega \tau)\ud \bm{\uprho}\ud \tau.
   \end{multline}
   Equation~(\ref{spectrogram_specific_intensity}) establishes a rigorous connection between the spectrogram and the
   specific intensity that is often
   used~\cite{andriyash_optoheterodyne_2016,andriyash_application_2018,kondratev_application_2020,andriyash_simultaneous_2022}.
   In particular, we note that the spectrogram at $\omega$ is the sum of the Doppler contributions at both $+\omega$ and
   $-\omega$. In the next section, we derive the transport equation satisfied by the specific intensity in a scattering
   and absorbing medium, accounting for both inelastic scattering and inhomogeneities in the optical properties.

   \section{Inelastic and inhomogeneous radiative transfer equation}\label{rte}
   % ===============================================================
   
   The RTE is a convenient tool to describe multiple scattering of light, due to its ability to handle complex
   geometries, and different transport regimes from single to deep multiple scattering. Historically, the RTE was
   established based on a phenomenological energy balance~\cite{CHANDRASEKHAR-1950}. Its derivation from the wave
   equation (first principles) is also well-known, and requires the definition of the specific intensity in statistical
   optics ~[Eq.~(\ref{specific_intensity_full})], together with a theory of multiple
   scattering~\cite{APRESYAN-1996,carminati_principles_2021}. In order to account for the motion of scatterers in an
   ejecta, and for spatial and temporal statistical heterogeneities in its optical properties, a generalized form of RTE
   is needed. The purpose of this section is to derive this generalized RTE from first principles, following the
   approach in Ref.~\onlinecite{pierrat_transport_2008} to account for Doppler shifts, and drawing inspiration from
   Ref.~\onlinecite{hoskins_radiative_2018} to handle statistical inhomogeneities in the medium (space and
   time-dependent particle number density, size and velocity distribution).

   \subsection{Bethe-Salpeter equation and statistical quasi-homogeneity}
   % ....................................................................
 
   The specific intensity, introduced in Eq.~(\ref{specific_intensity_full}), is based on the field-field correlation
   function, which is known to obey the Bethe-Salpeter equation. The latter is the
   starting point to derive the generalized RTE. It can be derived from first principles using the multiple
   scattering theory~\cite{RYTOV-1989,SHENG-2006,carminati_principles_2021}. We simply recall its formulation here. In
   the multiple scattering regime, light depolarizes after propagation over a distance on the order of the scattering
   mean-free path~\cite{VYNCK-2014}. For this reason, we consider the scalar approximation. Thus the Bethe-Salpeter is
   given by
   \begin{multline}\label{bethe_salpeter}
      \bra {E}(\bm{r},t){E}(\bm{\uprho},\tau)\ket=\bra {E}(\bm{r},t)\ket\bra {E}(\bm{\uprho},
      \tau)\ket
      +\int \bra {G}(\bm{r},\bm{r}',t,t')\ket
   \\\times
      \bra {G}(\bm{\uprho},\bm{\uprho}',\tau,\tau')\ket
      \Gamma(\bm{r}',\bm{r}'',\bm{\uprho}',\bm{\uprho}'',t',t'',\tau',\tau'')
   \\\times
      \bra {E}(\bm{r}'',t''){E}(\bm{\uprho}'',\tau'')\ket
      \ud\bm{r}'\ud\bm{r}''\ud\bm{\uprho}'\ud\bm{\uprho}''\ud t'\ud t''\ud\tau'\ud\tau''.
   \end{multline}
   We note that the field-field correlation function is written for the real field $E$ here while the specific intensity
   involved the analytic field. We will rigorously make the switch to $\bar{E}$ at the appropriate moment. Also,
   Eq.~(\ref{bethe_salpeter}) is written in the time domain, a formulation which is required to take into account the
   quasi-homogeneous aspect of the statistical properties of the medium. In this equation, $\bra G\ket$ is the average
   Green’s function. Physically, $\bra G (\bm{r},\bm{r}’,t,t’)\ket$ corresponds to the average electric field generated
   at point $\bm{r}$ and time $t$ by a point source at point $\bm{r}’$ emitting an infinitely short pulse at time $t’$.
   We note that $\bra G \ket$ is not translationally invariant in a statistically inhomogeneous medium. $\Gamma$ is known
   as the irreducible vertex, that contains the contributions of all scattering sequences connecting the field-field
   correlations functions at different points and times. The Bethe-Salpeter equation can therefore be seen as a propagation
   equation for the field-field correlation function.

   To derive the quasi-homogeneous RTE, we first assume that the statistical properties of the medium (typically the
   number density of particles, their size or their velocity distribution) are expected to vary slowly in space and time
   compared to all other characteristic lengths and times. We perform a change of variables first in
   the average Green function that reads
   \begin{equation}
      \bra {G}(\bm{r},\bm{r}',t,t')\ket=\bra \hat{G}\left(\bm{r}_0,\bm{r}-\bm{r}',t_0,t-t'\right)\ket
   \end{equation}
   where $\bm{r}_0=(\bm{r}+\bm{r}')/2$ and $t_0=(t+t')/2$ respectively.
   This change of variable is standard in coherence theory~\cite{MANDEL-1995}. In the following, $\bm{r}_0$ and
   $t_0$ will be kept fixed in particular when performing Fourier transforms thanks to the slow space and time evolution
   of the statistical properties of the medium. We perform a similar operation for the
   intensity vertex that becomes
   \begin{multline}
      \Gamma(\bm{r}',\bm{r}'',\bm{\uprho}',\bm{\uprho}'',t',t'',\tau',\tau'')
   \\
         =\hat{\Gamma}(\bm{r}_0,\bm{r}''-\bm{r}',\bm{\uprho}'-\bm{r}',\bm{\uprho}''-\bm{r}',t_0,t''-t',\tau'-t',\tau''-t').
   \end{multline}
   The Bethe-Salpeter equation can be rewritten in the form
   \begin{multline}\label{bethe_salpeter_centered}
      \bra {E}(\bm{r},t){E}(\bm{\uprho},\tau)\ket=\bra {E}(\bm{r},t)\ket\bra {E}(\bm{\uprho},\tau)\ket
   \\   
      +\int \bra \hat{{G}}\left(\bm{r}_0,\bm{r}-\bm{r}',t_0,t-t'\right)\ket
      \bra \hat{{G}}\left(\bm{r}_0,\bm{\rho}-\bm{\rho}',t_0,\tau-\tau'\right)\ket
   \\\times      
      \hat{\Gamma}\left(\bm{r}_0,\bm{r}''-\bm{r}',\bm{\uprho}'-\bm{r}',\bm{\uprho}''-\bm{r}',
      t_0,t''-t',\tau'-t',\tau''-t'\right)
   \\\times
      \bra {E}(\bm{r}'',t''){E}(\bm{\uprho}'',\tau'')\ket
      \ud\bm{r}'\ud\bm{r}''\ud\bm{\uprho}'\ud\bm{\uprho}''\ud t'\ud t''\ud\tau'\ud\tau''.
   \end{multline}
   This transformation in $\bra G \ket$ and $\Gamma$ is central in the so-called quasi-homogeneous approximation.
   Indeed, at fixed $\bm{r}_0$ and $t_0$, this amounts to considering the statistical properties of the medium
   as homogeneous. Taking the spatio-temporal Fourier transform of Eq.~(\ref{bethe_salpeter_centered}), leads to
   \begin{multline}\label{bethe_salpeter_centered_ft}
      \bra E(\bm{k},\omega)E^*(\bm{\kappa},\varpi)\ket=
      \int \bra \hat{{G}}\left(\bm{r}_0,\bm{k},t_0,\omega\right)\ket
      \bra \hat{{G}}^*\left(\bm{r}_0,\bm{\kappa},t_0,\varpi\right)\ket
   \\\times      
      \hat{\Gamma}\left(\bm{r}_0,\bm{k}',\bm{\kappa},\bm{\kappa}',
      t_0,\omega',\varpi,\varpi'\right)
      \bra E(\bm{k}',\omega')E^*(\bm{\kappa}',\varpi')\ket
   \\\times
      16\pi^2\delta(\bm{k}-\bm{k}'-\bm{\kappa}+\bm{\kappa}')\delta(\omega-\omega'-\varpi+\varpi')
   \\\times
      \frac{\ud\bm{k}'}{8\pi^3}\frac{\ud\bm{\kappa}'}{8\pi^3}\frac{\ud\omega'}{2\pi}\frac{\ud\varpi'}{2\pi}
   \end{multline}
   where we have dropped the product of average fields in the right-hand side that would contribute as a source term.
   The next step of the derivation consists in performing a second
   change of variables with $\bm{q} = \bm{k}-\bm{\kappa}$, $\bm{k} \rightarrow \bm{k}+\bm{q}/2$,
   $\bm{\kappa} \rightarrow \bm{k}-\bm{q}/2$, $\Omega =\omega-\varpi$, $\omega \rightarrow \omega+\Omega/2$
   and $\varpi \rightarrow \omega-\Omega/2$. Thus, Eq.~(\ref{bethe_salpeter_centered_ft})
   becomes
   \begin{multline}\label{bethe_salpeter_analytic_field}
      F(\bm{q},\bm{k},\Omega,\omega)=
      \\\int \bra \hat{G}\left(\bm{r}_0,\bm{k}+\frac{\bm{q}}{2},t_0,\omega+\frac{\Omega}{2}\right)\ket
      \bra \hat{G}^*\left(\bm{r}_0,\bm{k}-\frac{\bm{q}}{2},t_0,\omega-\frac{\Omega}{2}\right)\ket
   \\\times      
      \hat{\Gamma}\left(\bm{r}_0,\bm{k}'+\frac{\bm{q}}{2},\bm{k}-\frac{\bm{q}}{2},\bm{k}'-\frac{\bm{q}}{2},t_0,
      \omega'+\frac{\Omega}{2},\omega-\frac{\Omega}{2},\omega'-\frac{\Omega}{2}\right)
   \\\times 
      F(\bm{q},\bm{k}',\Omega,\omega')\frac{\ud\bm{k}'}{8\pi^3}\frac{\ud\omega'}{2\pi}
   \end{multline}
   where
   \begin{equation}\label{f_function}
      F(\bm{q},\bm{k},\Omega,\omega)=\bra E\left(\bm{k}+\frac{\bm{q}}{2},\omega+\frac{\Omega}{2}\right)E^*\left(\bm{k}
      -\frac{\bm{q}}{2},\omega-\frac{\Omega}{2}\right)\ket.
   \end{equation}
   To go further, the average Green function $\bra\hat{G}\ket$ needs to be explicitly given. It is driven
   by the Dyson equation that reads~\cite{DYSON-1949,DYSON-1949-1,RYTOV-1989,carminati_principles_2021}
   \begin{multline}\label{dyson}
      \bra {G}(\bm{r},\bm{r}',t,t')\ket={G}_0(\bm{r}-\bm{r}',t-t')+\int {G}_0(\bm{r}-\bm{r}'',t-t'')
   \\\times
      \Sigma(\bm{r}'',\bm{r}''',t'',t''')\bra {G}(\bm{r}''',\bm{r}',t''',t')\ket
      \ud\bm{r}''\ud\bm{r}'''\ud t''\ud t'''.
   \end{multline}
   The self energy $\Sigma$ is the operator that accounts for all the possible scattering sequences
   connecting the average field at different points and times. $G_0$ is the Green function for the homogeneous medium
   without the scattering particles (\ie, air). It corresponds to the electric field solution of the wave equation
   with a Dirac delta source term, and an outgoing radiation condition. It is given by
   \begin{equation}\label{green_air}
      \left[-\Delta_{\bm{r}}+\frac{1}{c^2}\frac{\partial^2}{\partial t^2}\right]
         G_0(\bm{r}-\bm{r}',t-t')=\delta(\bm{r}-\bm{r}')\delta(t-t').
   \end{equation}
   We perform again the change of variables defining a central position $\bm{r}_0$ and a central time $t_0$.
   Equation~(\ref{dyson}) becomes
   \begin{multline}\label{centered_dyson}
      \bra \hat{{G}}(\bm{r}_0,\bm{r}-\bm{r}', t_0,t-t')\ket={G}_0(\bm{r}-\bm{r}',t-t')
   \\
      +\int {G}_0(\bm{r}-\bm{r}'',t-t'')
      \hat{\Sigma}(\bm{r}_0,\bm{r}''-\bm{r}''',t_0,t-t'')
   \\\times
      \bra \hat{{G}}(\bm{r}_0,\bm{r}'''-\bm{r}',t_0,t''-t')\ket
      \ud\bm{r}''\ud\bm{r}'''\ud t''\ud t'''.
   \end{multline}
   To solve Eq.~(\ref{centered_dyson}), we take its Fourier transform and knowing that
   $G_0(\bm{k},\omega)=(k^2-\omega^2/c^2)^{-1}$, we get
   \begin{equation}\label{average_green}
      \bra \hat{G}(\bm{r}_0,\bm{k},t_0,\omega)\ket
      =\frac{1}{k^2-\omega^2/c^2-\hat{\Sigma}(\bm{r}_0,\bm{k},t_0,\omega)}.
   \end{equation}
   We now replace the average Green function by its expression given by Eq.~(\ref{average_green}) in
   Eq.~(\ref{bethe_salpeter_analytic_field}). Making use of the relation $(AB)^{-1}=(B-A)^{-1}(A^{-1}-B^{-1})$ with
   $A^{-1}=\bra \hat{G}\ket$ and $B^{-1}=\bra \hat{G}^*\ket$, we get
   \begin{widetext}
      \begin{multline}\label{exact_bethe_salpeter}
         \left[\frac{2\omega\Omega}{c^2}-2\bm{k}\cdot\bm{q}+\hat{\Sigma}\left(\bm{r}_0,\bm{k}+\frac{\bm{q}}{2},t_0,\omega
         +\frac{\Omega}{2}\right)
         -\hat{\Sigma}^*\left(\bm{r}_0,\bm{k}-\frac{\bm{q}}{2},t_0,\omega-\frac{\Omega}{2}\right)\right]
         F(\bm{q},\bm{k},\Omega,\omega)
      \\ =
         \left[\bra\hat{G}\left(\bm{r}_0,\bm{k}+\frac{\bm{q}}{2},t_0,\omega+\frac{\Omega}{2}\right)\ket
         -\bra\hat{G}^*\left(\bm{r}_0,\bm{k}-\frac{\bm{q}}{2},t_0,\omega-\frac{\Omega}{2}\right)\ket\right]
         \int\hat{\Gamma}\left(\bm{r}_0,\bm{k}'+\frac{\bm{q}}{2},\bm{k}-\frac{\bm{q}}{2},\bm{k}'-\frac{\bm{q}}{2},t_0,
         \omega'+\frac{\Omega}{2},\omega-\frac{\Omega}{2},\omega'-\frac{\Omega}{2}\right)
      \\\times
         F(\bm{q},\bm{k}'\Omega,\omega')\frac{\ud\bm{k}'}{8\pi^3}\frac{\ud \omega'}{2\pi}.
      \end{multline}
   \end{widetext}
   At this point, we have only performed an approximation of quasi-homogeneity of the statistical properties of the
   medium and additional approximations are required in order to obtain the RTE.

   \subsection{Weak extinction limit and radiative transfer equation}
   % ................................................................
   
   We now perform the large-scale approximation $|\bm{q}|\ll\{|\bm{k}|,|\bm{k'}|\}$ which is also known as the radiative
   transfer limit~\cite{BARABANENKOV-1968,RYZHIK-1996,carminati_principles_2021}. It assumes that the field-field
   correlation function varies on length scales $2\pi/|\bm{q}|$ much larger than the effective wavelength in the medium
   $2\pi/k_R$, the expression of which will be given below. Since the typical distance over which the field-field
   correlation function varies is the extinction mean-free path $\ell_e$, the large-scale limit holds as soon as
   $k_R\ell_e\gg 1$. This regime of weak extinction is relevant to ejecta in shock experiments.
   A large scale asymptotics for the time dependence is also relevant, and we assume
   $|\Omega|\ll\{|\omega|,|\omega'|\}$. Assuming non-resonant scatterers and weak non-locality, the $\bm{q}$ and
   $\Omega$ dependencies can be dropped in the average Green function $\bra \hat{G}\ket$, the self-energy $\hat{\Sigma}$
   and the intensity vertex $\hat{\Gamma}$ in Eq.~(\ref{exact_bethe_salpeter}). However, in order to obtain the correct
   expression for the transport velocity, it is crucial to keep the first term of the Taylor series in $\Omega$ for
   $\Re\hat{\Sigma}$. We get
   \begin{multline}\label{quasi_rte}
      \left[\frac{\omega\Omega}{c^2}-\bm{k}\cdot\bm{q}+i\Im\hat{\Sigma}\left(\bm{r}_0,\bm{k},t_0,\omega\right)
      +\frac{\Omega}{2}\Re\frac{\partial\hat{\Sigma}\left(\bm{r}_0,\bm{k},t_0,\omega\right)}{\partial \omega}\right]
   \\\times
      F(\bm{q},\bm{k},\Omega,\omega)=i\Im\bra\hat{G}\left(\bm{r}_0,\bm{k},t_0,\omega\right)\ket
   \\\times
      \int\hat{\Gamma}\left(\bm{r}_0,\bm{k}',\bm{k},\bm{k}',t_0,\omega',\omega,\omega'\right)
      F(\bm{q},\bm{k}',\Omega,\omega')\frac{\ud\bm{k}'}{8\pi^3}\frac{\ud \omega'}{2\pi}.
   \end{multline}
   The weak extinction regime also amounts to assuming $|\hat{\Sigma}|\ll k_0^2$. Thus the average Green function $\bra\hat{G}\ket$ in Eq.~(\ref{average_green}) is very peaked around
   $|\bm{k}|=k_0$~\cite{carminati_principles_2021}. In this case, the self-energy $\hat{\Sigma}$ can be evaluated on-shell
   at $|\bm{k}|=k_0$. By making use of the identity
   \begin{equation}
      \lim_{\epsilon\to 0^+}\frac{1}{x-x_0-i\epsilon}=\operatorname{PV}\left[\frac{1}{x-x_0}\right]+i\pi\delta(x-x_0),
   \end{equation}
   where $\operatorname{PV}$ is the Cauchy principal value, we find that
   \begin{equation}
      \Im\bra\hat{G}\left(\bm{r}_0,\bm{k},t_0,\omega\right)\ket
         =\pi\delta\left[k^2-k_0^2-\Re\hat{\Sigma}\left(\bm{r}_0,k_0,t_0,\omega\right)\right]
   \end{equation}
   where we have assumed that the disorder is statistically isotropic which means that $\hat{\Sigma}$ depends only on
   $k_0$. The relation above fixes the modulus of the wavevector $k=k_R(\bm{r}_0,t_0,\omega)$,
   with $k_R(\bm{r}_0,t_0,\omega)=\sqrt{k_0^2+\Re\hat{\Sigma}\left(\bm{r}_0,k_0,t_0,\omega\right)}$.
   We now introduce the group velocity as $1/v_g(\bm{r}_0,t_0,\omega)=\ud k_R(\bm{r}_0,t_0,\omega)/\ud \omega$
   and one can see that
   \begin{equation}\label{energy_velocity}
      \frac{\omega}{c^2}+\frac{1}{2}\Re\frac{\partial\hat{\Sigma}\left(\bm{r}_0,\bm{k},t_0,\omega\right)}{\partial \omega}
      =\frac{k_R(\bm{r}_0,t_0,\omega)}{v_g(\bm{r}_0,t_0,\omega)}.
   \end{equation}
   We note that for non-resonant scatterers the phase velocity $v_p$ equals the group velocity $v_g$ which also equals
   the transport (or energy) velocity $v_E$. We thus have $v_E=v_g=v_p=c/n_R$ where $k_R=n_R\omega/c$, $n_R$ being the
   real part of the effective refractive index of the medium. This leads to
   \begin{multline}\label{quasi_rte_bis}\hspace{-0.5cm}
      \left[\frac{\Omega k_R(\bm{r}_0,t_0,\omega)}{v_E(\bm{r}_0,t_0,\omega)}-k_R\bm{u}\cdot\bm{q}+i\Im\hat{\Sigma}\left(\bm{r}_0,k_R,t_0,\omega\right)
      \right]
      F(\bm{q},k_R\bm{u},\Omega,\omega)
   \\
      =i\pi\delta(k^2-k_R^2)
      \int\hat{\Gamma}\left(\bm{r}_0,k_R\bm{u}',k_R\bm{u},k_R\bm{u}',t_0,\omega',\omega,\omega'\right)
   \\\times
      F(\bm{q},k_R\bm{u}',\Omega,\omega')\frac{\ud\bm{k}'}{8\pi^3}\frac{\ud \omega'}{2\pi}.
   \end{multline}
   At this stage, it is important to note that the field entering the definition of the specific intensity [see
   Eq.~(\ref{specific_intensity_full})] is the analytic field $\bar{E}$, whereas the field entering the expression of
   our current field-field correlation $F$ [see Eq.~(\ref{f_function})] is the real field $E$. Since $\omega$ and
   $\omega'$ remain close to the incident frequency $\pm\omega_0$ thanks to a weak Doppler shift at each scattering
   event and thanks to the assumption $|\Omega|\ll\{|\omega|,|\omega'|\}$, the field-field correlation $F$ has two peaks
   in the frequency domains (one for positive frequencies and the other for negative ones) that do not overlap. Thus we
   can easily keep only positive frequencies which means that the real field can be replaced by the analytic field
   $\bar{E}$ in the $F$ function. Using the definition of the specific intensity given by
   Eq.~(\ref{specific_intensity_full}) and taking the inverse Fourier transform of Eq.~(\ref{quasi_rte_bis}) with
   respect to $\bm{q}$ and $\Omega$, we finally get
   \begin{multline}\hspace{-0.5cm}
      \left[\frac{1}{v_E(\bm{r}_0,t_0,\omega)}\frac{\partial}{\partial t}+\bm{u}\cdot\bm{\nabla}_{\bm{r}}
      +\frac{\Im\hat{\Sigma}\left(\bm{r}_0,k_R\bm{u},t_0,\omega\right)}{k_R}\right]
      I(\bm{r},\bm{u},t,\omega)
   \\
      =\frac{1}{16\pi^2}\int\hat{\Gamma}\left(\bm{r}_0,k_R\bm{u}',k_R\bm{u},k_R\bm{u}',t_0,\omega',\omega,\omega'\right)
   \\\times
      I(\bm{r},\bm{u}',t,\omega')\ud\bm{u}'\frac{\ud \omega'}{2\pi}
   \end{multline}
   where $\ud\bm{u}'$ means integration over the solid angle.  In the above equation, $\bm{r}_0$ and $t_0$ are slow variables
   compared to $\bm{r}$ and $t$, as a consequence of the quasi-homogeneous approximation. The equation is not modified if
   $\bm{r}_0$ and $t_0$ are replaced by $\bm{r}$ and $t$, respectively.
   The expression of the RTE in the quasi-homogeneous approximation and in the presence of inelastic scattering finally
   takes the form
   \begin{multline}\label{quasi_homogeneous_rte}
      \left[\frac{1}{v_E(\bm{r},t,\omega)}\frac{\partial}{\partial t}+\bm{u}\cdot\bm{\nabla}_{\bm{r}}+\frac{1}{\ell_e(\bm{r},
      t,\omega)}\right]
      I(\bm{r},\bm{u},t,\omega)
   \\
      =\frac{1}{\ell_s(\bm{r},t,\omega)}\int p(\bm{r},\bm{u},\bm{u}',t,\omega,\omega')
      I(\bm{r},\bm{u}',t,\omega')\ud\bm{u}'\frac{\ud \omega'}{2\pi}.
   \end{multline}
   In this equation the extinction mean-free path and the scattering mean-free path are respectively 
   given by
   \begin{align}
      \frac{1}{\ell_e(\bm{r},t,\omega)} & = \frac{\Im\hat{\Sigma}\left(\bm{r},k_R\bm{u},t,\omega\right)}{k_R},
   \\
      \frac{1}{\ell_s(\bm{r},t,\omega)} & =\frac{1}{16\pi^2} \int 
      \hat{\Gamma}\left(\bm{r},k_R\bm{u}',k_R\bm{u},k_R\bm{u}',t,\omega',\omega,\omega'\right)
      \ud\bm{u}'\frac{\ud\omega'}{2\pi}
   \end{align}
   while the phase function is given by
   \begin{multline}
      p(\bm{r},\bm{u},\bm{u}',t,\omega,\omega') =\frac{\ell_s(\bm{r},t,\omega)}{16\pi^2}
   \\\times
      \hat{\Gamma}\left(\bm{r},k_R\bm{u}',k_R\bm{u},k_R\bm{u}',t,\omega',\omega,\omega'\right).
      \label{eq:phase_function_general}
   \end{multline}
   We note that with these definitions the phase function is normalized as
   $\int p(\bm{r},\bm{u},\bm{u}',t,\omega,\omega')\ud\bm{u}'\ud\bm{\omega}'/(2\pi)=1$.
   
   Equation~(\ref{quasi_homogeneous_rte}), as the standard RTE, can be understood as an energy balance. 
   The first two terms in the left-hand side correspond to the spatio-temporal evolution of the specific intensity,
   which itself can be seen as a local (in space and time) and directional radiative flux at frequency $\omega$. 
   The third term describes losses by extinction (absorption and scattering). The right hand side
   corresponds to gain by scattering. The phase function describes the fraction of the incoming power along
   direction $\bm{u}'$ and at frequency $\omega'$ that is scattered along direction $\bm{u}$ at frequency $\omega$,
   with the change of frequency corresponding to the Doppler shift acquired during the inelastic scattering process.

   \subsection{Practical expressions of mean-free paths and phase function}
   % ----------------------------------------------------------------------

   In order to derive explicit expressions for the mean-free paths and the phase function, we need to start from the
   expression for the $t$-matrix (or scattering matrix) of an individual scatterer. In Fourier space, and for a particle
   with velocity $\bm{v}$, it can be written as~\cite{pierrat_transport_2008}
   \begin{multline}\label{scattering_operator}
      t_{\bm{v}}(a,\bm{k},\bm{k}',\omega,\omega')=2\pi t(a,\bm{k},\bm{k}',\omega-\bm{k}\cdot\bm{v})
   \\\times
      \delta[\omega'-\omega-(\bm{k}'-\bm{k})\cdot\bm{v}]
   \end{multline}
   with $t(a,\bm{k},\bm{k}',\omega)$ the $t$-matrix of a static particle with radius $a$. The Dirac delta function accounts for
   the Doppler shift. Given the velocities involved, the condition $|\bm{k}\cdot\bm{v}|\ll|\omega|$ is satisfied, which 
   implies that the shift can be neglected in the $t$-matrix, the latter being a slowly varying function of frequency.
   
   The self-energy is a rather complex object and we need some approximations to go further. In the weak-extinction regime, 
   we can use the lowest order approximation to $\Sigma$, know as the independent scattering approximation (ISA). This leads 
   to
   \begin{multline}\label{sigma}
      \Sigma(\bm{r},\bm{r}',t,t')=\sum_{i=1}^{N(t)} \int t_{\bm{v}_i}(a_i,\bm{r}-\bm{r}_i,\bm{r}'-\bm{r}_i,t,t')
   \\\times
      P(\bm{r}_i,t,a_i,\bm{v}_i)\ud\bm{r}_i \ud a_i \ud\bm{v}_i
   \end{multline}
   where $P(\bm{r}_i,t,a_i,\bm{v}_i)$ is the probability density at time $t$ of finding particle $i$ at position
   $\bm{r}_i$, with a radius $a_i$ and a velocity $\bm{v_i}$. Note that this expression is very general, but in order to
   have the particle number density $\rho(\bm{r}_i,t)$ appearing explicitly we performed the following splitting.  We
   assume that $P(\bm{r}_i,t,a_i,\bm{v}_i)=\rho(\bm{r}_i,t)g(\bm{r}_i,t,a_i,\bm{v}_i)/N(t)$, where
   $g(\bm{r}_i,t,a_i,\bm{v}_i)$ is now the probability density at time $t$ and position $r_i$ of finding particle $i$
   with a radius $a_i$ and a velocity $\bm{v}_i$.  From this expression, using again the quasi-homogeneous approximation
   the self-energy in Eq.~(\ref{sigma}) can be rewritten as
   \begin{multline}\label{sigma_centered}
      \hat{\Sigma}(\bm{r}_0,\bm{r}-\bm{r}',t_0,t-t')=
      \rho(\bm{r}_0,t_0)
   \\\times
      \int t_{\bm{v}_i}(a_i,\bm{r}-\bm{r}_i,\bm{r}'-\bm{r}_i,t,t')
   \\\times
      g(\bm{r}_0,t_0,a_i,\bm{v}_i)\ud\bm{r}_i\ud a_i\ud\bm{v}_i.
   \end{multline}
   Taking the Fourier transform of Eq.~(\ref{sigma_centered}) and making use of Eq.~(\ref{scattering_operator}), we finally 
   get
   \begin{equation}\label{sigma_quasi_homogene}
      \hat{\Sigma}(\bm{r}_0,k_R\bm{u},t_0,\omega)=
      \rho(\bm{r}_0,t_0)\int t(a,k_R\bm{u},k_R\bm{u},\omega)
      h(\bm{r}_0,t_0,a)\ud a.
   \end{equation}
   In this expression $h(\bm{r}_0,t_0,a)=\int g(\bm{r}_0,t_0,a,\bm{v})\ud\bm{v}$ is the probability density of having a
   particle with radius $a$, knowing that it lies at position $\bm{r}_0$ at time $t_0$. We note that the Doppler shift
   does not enter the expression of the self-energy. This can be easily understood: The self-energy defines the average
   field which propagates in the forward direction, as can be seen in the $(\bm{k},\bm{k})$ dependence of the $t$-matrix
   in Eq.~(\ref{sigma_quasi_homogene}), for which there is no Doppler shift.  Finally, the expression of the extinction
   mean-free path is
   \begin{equation}
      \frac{1}{\ell_e(\bm{r},t,\omega)}
         =\rho(\bm{r},t)\int \sigma_{e}(a,\omega)h(\bm{r},t,a)\ud a
   \end{equation}
   where $\sigma_{e}(a,\omega)=\Im t(a,k_R\bm{u},k_R\bm{u},\omega)/k_R$ is the extinction cross-section of a particle 
   with radius $a$. 
   
   To compute the phase function, we need to express the intensity vertex. The derivation is identical to that performed
   for the self-energy. In the same weak-scattering limit, the intensity
   vertex $\Gamma$ is given by
   \begin{multline}\label{intensity_vertex}
      \Gamma(\bm{r},\bm{r}',\bm{\uprho},\bm{\uprho}',t,t',\tau,\tau'')
   \\=
      \sum_{i=1}^{N(t)}  \int t_{\bm{v}_i}(a_i,\bm{r}-\bm{r}_i,\bm{r}'-\bm{r}_i,t,t')
   \\\times
      t_{\bm{v}_i}(a_i,\bm{\rho}-\bm{r}_i,\bm{\rho}'-\bm{r}_i,\tau,\tau')
   \\\times
      P(\bm{r}_i,t,a_i,\bm{v}_i)\ud\bm{r}_i \ud a_i \ud\bm{v}_i.
   \end{multline}
   In particular, from Eqs.~(\ref{intensity_vertex}) and (\ref{scattering_operator}), we obtain
   \begin{multline}
      \hat{\Gamma}\left(\bm{r},k_R\bm{u}',k_R\bm{u},k_R\bm{u}',t,\omega',\omega,\omega'\right)
      =\rho(\bm{r},t)
   \\\times
      \int
      \left|t(a,k_R\bm{u},k_R\bm{u}',\omega)\right|^2
      2\pi\delta\left[\omega'-\omega-k_R(\bm{u}'-\bm{u})\cdot\bm{v}\right]
   \\\times
      g(\bm{r},t,a,\bm{v})\ud a\ud\bm{v}.
   \end{multline}
   In this expression, the Doppler shift appears explicitly. We also note that the square modulus of the
   $t$-matrix is proportional to the differential scattering cross-section of a particle through the relation
   ~\cite{carminati_principles_2021}
   $\ud\sigma_{s}(a,\bm{u}\cdot\bm{u}',\omega)/\ud\bm{u}=\left|t(a,k_R\bm{u},k_R\bm{u}',\omega)\right|^2/(16\pi^2)$.
   Inserting the above equation in Eq.~(\ref{eq:phase_function_general}) allows us to obtain the explicit expression of the 
   phase function
   \begin{multline}
      \frac{1}{\ell_s(\bm{r},t,\omega)} p(\bm{r},\bm{u},\bm{u}',t,\omega,\omega')
         =\rho(\bm{r},t)\int \frac{\ud\sigma_{s}(a,\bm{u}\cdot\bm{u}',\omega)}{\ud\bm{u}}
   \\\times
         2\pi\delta\left[\omega'-\omega-k_R(\bm{u}'-\bm{u})\cdot\bm{v}\right]
         g(\bm{r},t,a,\bm{v})\ud a\ud\bm{v}.
   \end{multline}
   Since we consider spherical particles for the sake of simplicity, the differential scattering cross-section and the phase
   function only depends on $\bm{u}\cdot\bm{u}'$, or equivalently on the angle between $\bm{u}'$ and $\bm{u}$.

   \section{Numerical simulations}\label{numerics}
   % =============================

   The quasi-homogeneous and inelastic RTE given by Eq.~(\ref{quasi_homogeneous_rte}) cannot be solved analytically
   in real geometries such as that corresponding to a shock ejecta. Monte-Carlo algorithms have become powerful and
   convenient tools to solve the RTE numerically, especially thanks to their simplicity of implementation and to the 
   ever-increasing available computer resources~\cite{SIEGEL-1992}. In this section we provide a justification of the
   correspondence between the output of a Monte-Carlo simulation and the solution to the generalized RTE.

   \subsection{Integral form of the RTE}
   % -----------------------------------

   We start by writing the solution to the RTE in the form of an integral equation, suitable to be solved by a Monte-Carlo
   algorithm. To proceed, we note that the time derivative in Eq.~(\ref{quasi_homogeneous_rte}) can be neglected 
   since the illumination in the experiment is monochromatic (steady-state) and the transit time for light withing the 
   scattering cloud remains short compared to the evolution time scale of the cloud $T_c$. In this case,
   Eq.~(\ref{quasi_homogeneous_rte}) simplifies into
   \begin{multline}
      \left[\bm{u}\cdot\bm{\nabla}_{\bm{r}}+\frac{1}{\ell_e(\bm{r},t,\omega)}\right]
         I(\bm{r},\bm{u},t,\omega)
         =\frac{1}{\ell_s(\bm{r},t,\omega)}
   \\\times
         \int p(\bm{r},\bm{u},\bm{u}',t,\omega,\omega')
            I(\bm{r},\bm{u}',t,\omega')\ud\bm{u}'\frac{\ud \omega'}{2\pi}.
            \label{eq:RTE_MC}
   \end{multline}
   We see that the time variable $t$ now plays the role of a parameter. In practice, this means that a steady-state 
   Monte-Carlo simulation will be performed for each time $t$ in the evolution of the spectrogram. 
   Considering the right-hand side in Eq.~(\ref{eq:RTE_MC}) as a source term for the RTE, denoted by
   $S(\bm{r},\bm{u},t,\omega)$ in the following, we first define the Green function $I_0$ as the solution to
   \begin{equation}
      \left[\bm{u}\cdot\bm{\nabla}_{\bm{r}}+\frac{1}{\ell_e(\bm{r},t,\omega)}\right]
         I_0(\bm{r},\bm{r}_0,\bm{u},t,\omega)
         =\delta(\bm{r}-\bm{r}_0)
   \end{equation}
   with the boundary condition $I_0(\bm{r},\bm{r}_0,\bm{u},t,\omega)\to 0$ when
   $|\bm{r}|\to\infty$. Defining $r_{\parallel}=\bm{r}\cdot\bm{u}$ and $\bm{r}_{\perp}=\bm{r}-(\bm{r}\cdot\bm{u})\bm{u}$,
   it is easy to show that
   \begin{multline}
      I_0(\bm{r},\bm{r}_0,\bm{u},t,\omega)
         =\delta(\bm{r}_{\perp}-\bm{r}_{0,\perp})\operatorname{H}(r_{\parallel}-r_{0,\parallel})
   \\\times
         \exp\left[-\int_{r_{0,\parallel}}^{r_{\parallel}}\frac{\ud s'}{\ell_e(\bm{r}_{0,\perp}+s'\bm{u},t,\omega)}\right]
   \end{multline}
   where the splitting
   $\delta(\bm{r}-\bm{r}_0)=\delta(\bm{r}_{\perp}-\bm{r}_{0,\perp})\delta(r_{\parallel}-r_{0,\parallel})$ has been used for
   the Dirac delta function, and $\operatorname{H}$ is the Heaviside step function. The solution to the RTE Eq.~(\ref{eq:RTE_MC})
   can now be written
    \begin{equation}
      I(\bm{r},\bm{u},t,\omega)=\int I_0(\bm{r},\bm{r}_0,\bm{u},t,\omega)S(\bm{r}_0,\bm{u},t,\omega)\ud\bm{r}_0
   \end{equation}
   which directly leads to
    \begin{multline}\label{integral_rte}
      I(\bm{r},\bm{u},t,\omega)
         =\int_{s=0}^{\infty}\frac{1}{\ell_s(\bm{r}-s\bm{u},t,\omega)}
         \exp\left[-\int_0^s\frac{\ud s'}{\ell_e(\bm{r}-s'\bm{u},t,\omega)}\right]
   \\\times
         \int_{\omega'}\int_{\bm{u}'}p(\bm{r}-s\bm{u},\bm{u},\bm{u}',t,\omega,\omega')
         I(\bm{r}-s\bm{u},\bm{u}',t,\omega')\ud\bm{u}'\frac{\ud \omega'}{2\pi}\ud s.
   \end{multline}
   This is the integral form of the quasi-homogeneous RTE, from which a Monte-Carlo scheme can be rigorously introduced.

   \subsection{Monte-Carlo scheme}
   % -----------------------------

   The Monte-Carlo scheme can be seen as a random walk process for energy quanta (behaving as classical particles),
   where each step is characterized by statistical distributions in step length, scattering direction, and frequency. 
   Equation~(\ref{integral_rte}) allows us to
   write explicitly the probability density $l$ of having a step of length $s$, in the form
   \begin{equation}\label{proba_density_length}
      l(\bm{r},t,s)=\frac{1}{\ell_s(\bm{r}-s\bm{u},t,\omega)}\exp\left[-\int_0^s\frac{\ud s'}{\ell_s(\bm{r}-s'\bm{u},t,
      \omega)}\right]
   \end{equation}
   where
   \begin{equation}
      \frac{1}{\ell_s(\bm{r},t,\omega)}
         =\rho(\bm{r},t)\int \sigma_{s}(a,\omega)h(\bm{r},t,a)\ud a,
   \end{equation}
   $\sigma_{s}(a,\omega)$ being the scattering cross section of a spherical particle with radius $a$. We note that the probability
   density $l$ is defined through the scattering mean-free path $\ell_s$ only. Absorption is taken into
   account through a weighting factor (for each energy quantum in the random walk) given by
   \begin{equation}
      w(\bm{r},t,s)=\exp\left[-\int_0^s\frac{\ud s'}{\ell_a(\bm{r}-s'\bm{u},t,\omega)}\right]
   \end{equation}
   where $\ell_a^{-1}=\ell_e^{-1}-\ell_s^{-1}$ is the absorption mean-free path.
   Similarly, the probability density of having scattered direction and frequency $(\bm{u},\omega)$ for
   incident direction and frequency $(\bm{u}',\omega')$ is
   \begin{multline}
      p(\bm{r},\bm{u},\bm{u}',t,\omega,\omega')
         =\left[\int \sigma_{s}(a,\omega)h(\bm{r},t,a)\ud a\right]^{-1}
   \\\times
         \int \frac{\ud\sigma_{s}(a,\bm{u}\cdot\bm{u}',\omega)}{\ud\bm{u}}
         2\pi\delta\left[\omega'-\omega-k_R(\bm{u}'-\bm{u})\cdot\bm{v}\right]
   \\\times
         g(\bm{r},t,a,\bm{v})\ud a\ud\bm{v} ,
   \end{multline}
   which is simply the phase function $p$ introduced previously. In practice, the optical properties
   of individual spherical particles, such as the extinction cross-section $\sigma_{e}$ and the differential scattering 
   cross-section $\ud\sigma_{s}/\ud\bm{u}$, are computed numerically using the Mie theory~\cite{MIE-1908}, with
   an average over the two possible incident polarization states for unpolarized light. In the shock ejecta geometry,
   we also have to take into account the free surface (which corresponds to the remaining bulk portion of the metallic 
   plate that was initially shocked). This is done by considering specular reflection at the interface to define a reflected 
   direction, followed by the use of the Doppler shift formula to define the reflected frequency.

   \subsection{Implementation}
   % -------------------------
      
   In the practical implementation of the Monte-Carlo simulation, we discretize space into several layers or cells,
   depending on space variations of the statistical properties of the particle in the cloud. In a given layer or cell, 
   the scattering and absorption mean-free paths $\ell_s$ and $\ell_a$, and the phase function $p$, are taken to be
   constant. They are computed thanks to the routine given in Ref.~\onlinecite{bohren_absorption_1983}.
   The probe geometry is mimicked by starting each random walk from the center of the probe, 
   with a direction along the optical axis and towards the scattering medium. The collection by the probe is represented
   by an acceptance cone with angle $\theta_p$ and the necessity for the energy quanta to hit a pupil with size $\phi_p$
   centered at the probe location.

   To draw a distance $s$ between two consecutive scattering events, we first normalize all lengths in each layer or
   cell by the corresponding scattering mean-free path $\ell_s$. Then, we randomly draw a distance $d$ in an exponential
   distribution with unit parameter, and rescale the distances with $\ell_s$ as we propagate from layer to layer or from
   cell to cell until we reach the end of the segment. To draw scattered direction and frequency, we first draw a
   velocity $\bm{v}$ and a particle radius $a$ following the distribution $g(\bm{r},t,a,\bm{v})$.  Then, we draw a
   scattered direction $\bm{u}$ from the differential scattering cross-section, and finally we compute the outgoing
   frequency $\omega'$ using the Dirac delta function. This process is certainly not optimized in terms of computation
   time, but remains low cost in term of computer memory.  Thus we are able to compute the detected flux $\int_G
   I(\bm{r},\bm{u},t,\omega_0+\omega)\bm{u}\cdot\bm{n}\ud\bm{u}\ud\bm{r}$ which, once symmetrised on $\omega$, gives the
   spectrogram at time $t$ thanks to Eq.~(\ref{spectrogram_specific_intensity}).  Repeated for different times spaced on
   the order of $T_c$, we reconstruct the full spectrogram $S(t,\omega)$.
   
   \subsection{Signature of multiple scattering}\label{multiple_scattering_regime}
   % -------------------------------------------

   In order to analyze the influence of multiple scattering on the spectrograms by Monte-Carlo simulations,
   we need to specify the parameters characterizing a realistic cloud of particles. We consider the simple scenario 
   used by Shi \emph{et al}~\cite{shi_reconstruction_2022}, and formulate the following assumptions: 
   (1) the ejection is along the $z$ direction, with particles propagating towards $z>0$, (2) the cloud is
   translationally invariant in the $x$ and $y$ direction, (3) the particle positions $z$ are directly given by $vt$, and
   (4) the particle radii and velocities are statistically independent, and do not depend on position and time,
   meaning that the probability density $g$ takes the form $g(\bm{r},t,a,\bm{v})=h(a)j(v)$.

   Regarding the particle sizes, we assume a lognormal distribution~\cite{schauer_ejected_2017}
   \begin{equation}
      h(a)=\frac{1}{a\sigma\sqrt{2\pi}}\exp\left[-\frac{(\ln a - \mu)^2}{2\sigma^2}\right],
   \end{equation}
   where $\mu=\ln(\mu_a^2/\sqrt{\mu_a^2+\sigma_a^2})$ and $\sigma^2=\ln(1+\sigma_a^2/\mu_a^2)$, $\mu_a$ and $\sigma_a$
   being, respectively, the mean and standard deviation of the particle radius $a$. For the velocity statistics, we assume
   a probability density
   \begin{equation}
      j(v)=\frac{\beta}{v_s}\exp\left[-\beta \left(\frac{v}{v_s}-1\right)\right]
   \end{equation}
   where the parameter $\beta$ gives the slope of the distribution, and $v_s$ is the velocity of the free 
   surface~\cite{monfared_experimental_2014}. From these two distributions, we deduce that the number density 
   of particles and the optical thickness take the form
   \begin{equation}
      \rho(\bm{r},t)=\frac{M_s}{\bar{V}\rho_{\text{Sn}}t}j\left(\frac{z}{t}\right),
   \end{equation}
   \begin{equation}\label{optical_thickness}
      b=\frac{M_s\bar{\sigma_s}}{\bar{V}\rho_{\text{Sn}}}
   \end{equation}
   where $\bar{V}=(4\pi/3)\int a^3h(a)\ud a$ and $\bar{\sigma_s}=\int\sigma_s(a)h(a)\ud a$ are the average volume
   and average scattering cross section of the particles, respectively, $\rho_{\text{Sn}}$ is the volumetric mass density 
   and $M_s$ is the total ejected mass per unit area. We note that the expression (\ref{optical_thickness}) of the optical
   thickness is similar to that appearing in earlier works
   ~\cite{andriyash_optoheterodyne_2016,andriyash_application_2018,andriyash_photon_2020,kondratev_application_2020,
   andriyash_simultaneous_2022},
   except that it is defined using the scattering mean-free path instead of the transport mean free path.
   We use the parameters~\cite{shi_reconstruction_2022} $\mu_a=\SI{0.75}{\micro m}$, $\sigma=0.5$, $M_s=\SI{20}{mg/cm^2}$
   and $\beta=10$, and we impose minimum and maximum values for the radii and the velocities 
   $a_{\text{min}}=\SI{0.1}{\micro m}$, $a_{\text{max}}=\SI{2.0}{\micro m}$, $v_{\text{min}}=v_s=\SI{2250}{m/s}$ and
   $v_{\text{max}}=\SI{4500}{m/s}$. With these parameters the optical thickness is $b=42$, showing that we consider
   the deep multiple scattering regime. 
   
   \begin{figure}[!htb]
      \centering
      \includegraphics[width=1.0\linewidth]{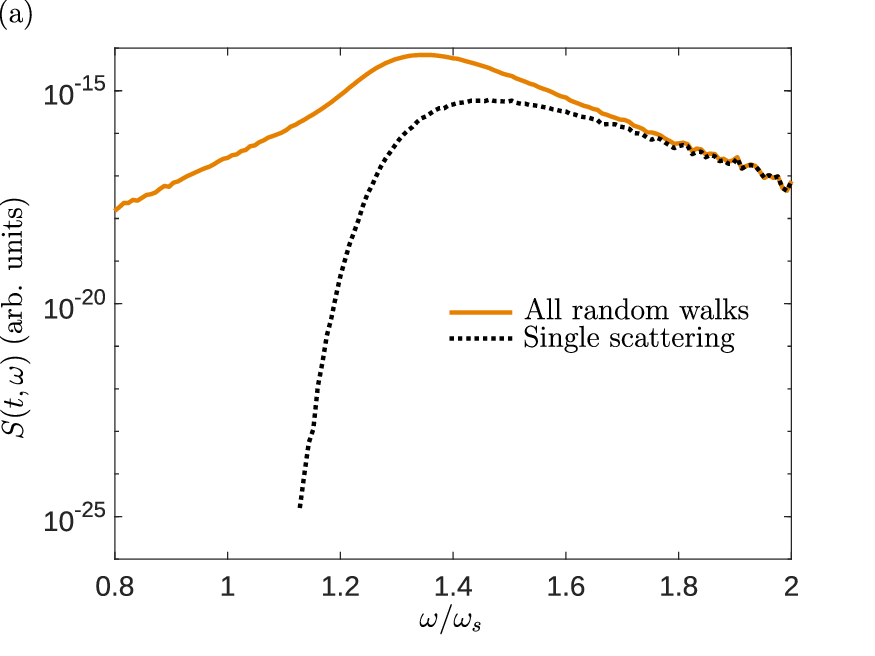}
      \includegraphics[width=1.0\linewidth]{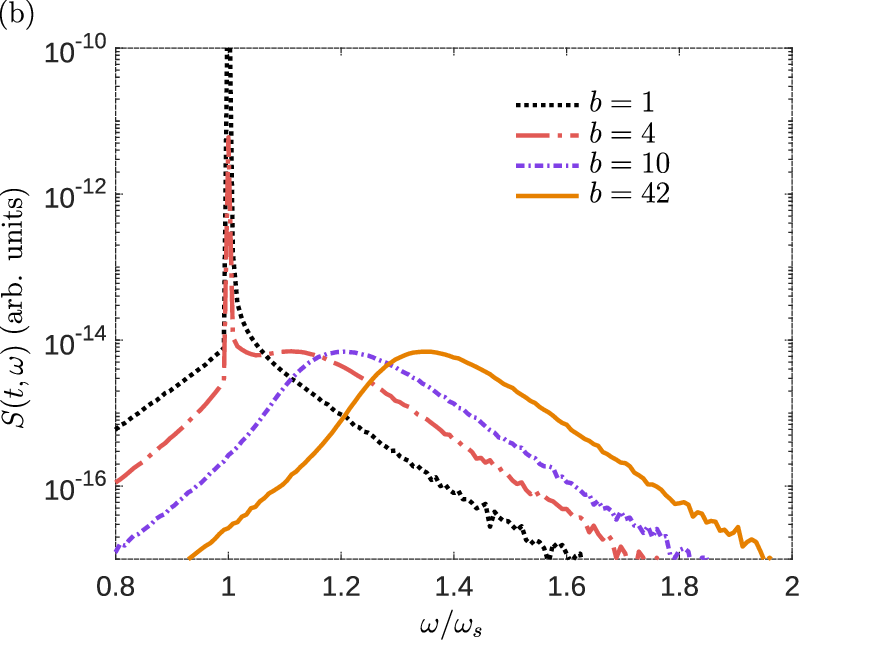}
      \caption{(a)~Comparison between single scattering (black dashed line) and multiple scattering spectra (red solid
      line). The simulations are carried out with the ejecta parameters
      defined in Sec.~\ref{multiple_scattering_regime}, and for a time $t=\SI{10}{\micro s}$. (b)~Spectra for different
      optical thicknesses. Same parameters as in (a). The variations in the optical thickness is obtained by changing 
      $M_s$, the total ejected mass per unit area. From left to right
      $M_s=\SI{0.47}{mg/cm^2}$, $M_s=\SI{1.41}{mg/cm^2}$, $M_s=\SI{4.71}{mg/cm^2}$ and $M_s=\SI{20}{mg/cm^2}$
      resulting in $b=1$, $b=4$, $b=10$ and $b=42$.}
      \label{shi}
   \end{figure}

   In Fig.~\ref{shi}, we present numerical simulations of spectrograms for a given time in the dynamics of the ejecta.
   Figure~\ref{shi}\,(a) compares a spectrum resulting from the full Monte-Carlo simulation (red solid line), and from a
   simulation restricted to single scattering events (black dashed line). The lineshape of the spectra are consistent
   with previously reported results~\cite{andriyash_optoheterodyne_2016,kondratev_application_2020}.  For high
   frequencies, the single scattering spectrum matches the spectrum computed with all multiple scattering sequences.
   This confirms that the front side of the ejecta scatterers chiefly in the single scattering regime, since the front
   side contains the fastest particles producing large Doppler shifts. For lower frequencies, as suggested by
   Franzkowiak~\emph{et~al.}~\cite{franzkowiak_multiple_2018} and
   Andriyash~\emph{et~al.}~\cite{andriyash_optoheterodyne_2016}, the signal is dominated by high-order scattering.  More
   specifically, for frequencies below the free-surface frequency $\omega_s$, the signal is only due to high-order
   scattering. This means that light that penetrates deep into the ejecta is inevitably multiply scattered.  We also
   observe that while multiple scattering contributions could have been expected to generate random frequencies, they
   actually contribute in the same range as the single scattering spectrum. On closer inspection, we find that the
   multiple scattering sequences consist mainly of a single backscattering event surrounded by forward scattering
   events. Since forward scattering produces low Doppler shifts, according to Eq.~(\ref{doppler}), the cumulated Doppler
   shifts of forward scattering events remains small compared to the shift produced by backscattering.  As a result,
   most multiple scattering sequences still encode the velocity of the particle they backscattered on, with an
   uncertainty produced by a broadening generated by the series of forward scattering events. The broadening remains
   small a the front side of the ejecta (where single scattering dominates) and becomes larger as light penetrates
   deeper into the particle cloud.
   
   Figure~\ref{shi}\,(b) compares spectra calculated for different optical thicknesses $b$ of the particle cloud.
   To change the optical thickness, we take the surface mass $M_s$ as a tunable parameter, keeping all other parameters
   constant. We see that a peak corresponding to the velocity of the free surface emerges for $b \leq 4$, and is not visible
   at larger optical thicknesses. This observation is consistent with results reported in previous
   works~\cite{andriyash_application_2018,kondratev_application_2020,shi_reconstruction_2022}.
   
   \subsection{Inhomogeneities in the particle size distributions}\label{size_distributions_inhomogeneities}
   % -------------------------------------------------------------

   The numerical simulations presented above included the full set of assumptions listed at the beginning of 
   Sec.~\ref{multiple_scattering_regime}, including homogeneity in the particle size-velocity distribution. 
   Molecular dynamics simulations~\cite{durand_large-scale_2012,durand_power_2013,durand_mass-velocity_2015} 
   and holography measurements~\cite{sorenson_ejecta_2014,schauer_ejected_2017,guildenbecher_ultraviolet_2023} 
   suggest that fragmentation tends to produce smaller particles at the beginning of the ejection process and larger 
   particles at the end. This results in a spatially inhomogeneous distribution of particle sizes. While the model
   based on Eq.~(\ref{quasi_homogeneous_rte}) and the associated Monte-Carlo simulations handle inhomogeneities
   in the particle density, it might be relevant to include inhomogeneities in the statistical distribution of particle size.
   The only comparable works are from Andriyash~\emph{et al.}~\cite{andriyash_photon_2020} and 
   Kondratev~\emph{et al.}~\cite{kondratev_application_2020}. Their model applies to an ejacta expanding in air,
   in which the slowing down of particles depends on their size. This also results in an
   ejecta with an inhomogeneous size distribution, but of different nature.

   In this work, we focus on the configuration studied by Sorenson~\emph{et al.}~\cite{sorenson_ejecta_2014}. 
   An ejecta is produced in vacuum in a shock experiment, and a holographic setup allows them to measure the 
   size distributions for five velocities in the ejecta. The experimental results confirm that the particles are larger 
   at slower velocities. To include this degree of freedom in our model, we reformulate the basic assumptions
   as follows: the ejection is along the $z$ direction, with particles propagating towards $z>0$, (2) the cloud is
    translationally invariant in the $x$ and $y$ direction, (3) the particle positions $z$ are directly given by $vt$,
   (4) the particle radii and velocities are statistically {\it dependent}, and do not depend on position
   and time, meaning that the probability density $g$ takes the form $g(\bm{r},t,a,\bm{v})=j(v)h(a,v)$.

   Regarding the statistics of particle radii, we keep a lognormal distribution~\cite{schauer_ejected_2017}
   \begin{equation}
      h(a,v)=\frac{1}{a\sigma(v)\sqrt{2\pi}}\exp\left\{-\frac{\left[\ln a - \mu(v)\right]^2}{2\sigma(v)^2}\right\},
   \end{equation}
   with parameter $\mu(v)$ and $\sigma(v)$ that now depend on the particle velocity. We fit these parameters 
   on the five experimental distributions given in Ref.~\onlinecite{sorenson_ejecta_2014}. 
   Then we extrapolate this set of five couples $(\mu_i,\sigma_i)$ by a fit of $\mu(v)$ and $\sigma(v)$ on all velocities. 
   This leads to $\mu(v)=\ln\left[\SI{3.6185e-6}\times(v/v_s)^{-1.487}\right]$ and
   $\sigma(v)=1.5180\times\exp\left[\mu(v)\right]+0.358$.

   From there, to retrieve $j(v)$, we use the measured mass-velocity distribution $M(v)$ defined as
   \begin{equation}
      M(v)=M_s\int_v^{+\infty}j(v')\frac{\bar{V}(v)}{\bar{V}}\ud v'
   \end{equation}
   where $\bar{V}(v)=\int(4/3)\pi a^3\rho_{\text{Sn}}h(a,v)\ud a$ is the average volume of particles at velocity $v$, and
   $\bar{V}=\int \bar{V}(v)j(v)\ud v$ is the average volume of particles in the entire ejecta. Experimentally $M(v)$
   follows an exponential law~\cite{monfared_experimental_2014} such that
   $M(v)=M_s\exp\left[-\beta\left(v/v_s-1\right)\right]$ with $M_s=\SI{35}{\milli g / \centi m^2}$,
   $\beta=4.789$ and $v_s=\SI{2250}{m/s}$. This gives
   \begin{equation}
      j(v)=\frac{\bar{V}}{\bar{V}(v)}\frac{\beta}{v_s}\exp\left[-\beta\left(\frac{v}{v_s}-1\right)\right].
   \end{equation}
   This expression is similar to that given in Sec.~\ref{multiple_scattering_regime}, except that here the factor
    $\bar{V}/\bar{V}(v)$ accounts for the inhomogeneity in the size distribution. As before, we can deduce the 
    number density of particles and the optical thickness, that take the form
   \begin{equation}
      \rho(\bm{r},t)=\frac{M_s}{\bar{V}\rho_{\text{Sn}}t}j\left(\frac{z}{t}\right),
   \end{equation}
   \begin{equation}
      b=\int\frac{M_s\bar{\sigma_s}(v)}{\bar{V}(v)\rho_{\text{Sn}}}j(v) \ud v
   \end{equation}
   where $\bar{\sigma_s}(v)= \int \sigma_s(a)h(a,v)\ud a$ is the average scattering cross section of the particles at 
   velocity $v$.

   Finally, to define a reference case for comparisons, we fit a lognormal size distribution over all particles in the
   ejecta. This describes an ejecta similar to that considered in Sec.~\ref{multiple_scattering_regime}, where
   inhomogeneities in the particle size distribution are not accounted for. In this case we get
   $\exp(\mu)=\SI{2.5750e-06}{m}$ and $\sigma=0.904$, while keeping the same values for $M_s$, $\beta$ and $v_s$.
   
   \begin{figure}[!htb]
      \centering
      \includegraphics[width=0.95\linewidth]{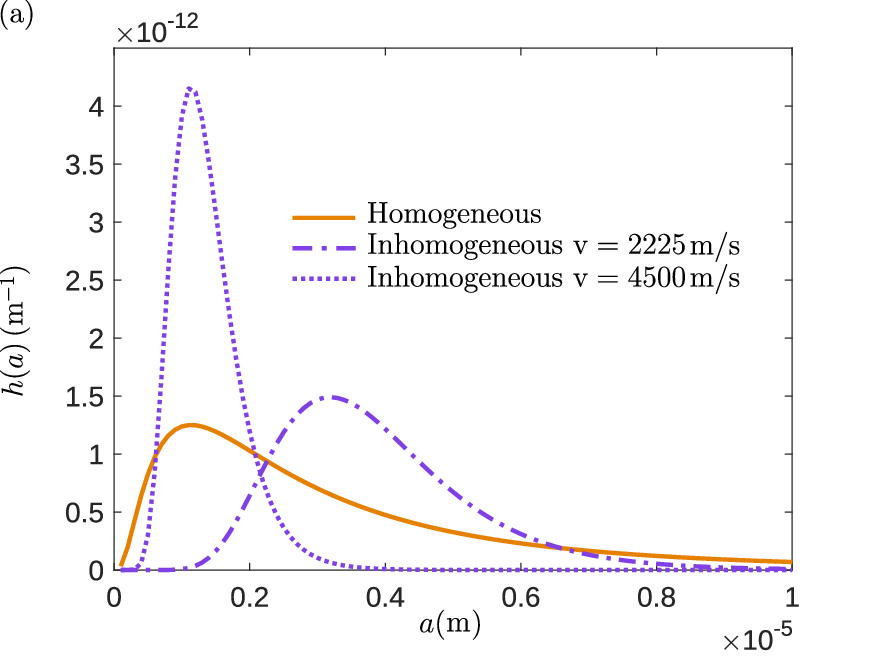}
      \includegraphics[width=0.95\linewidth]{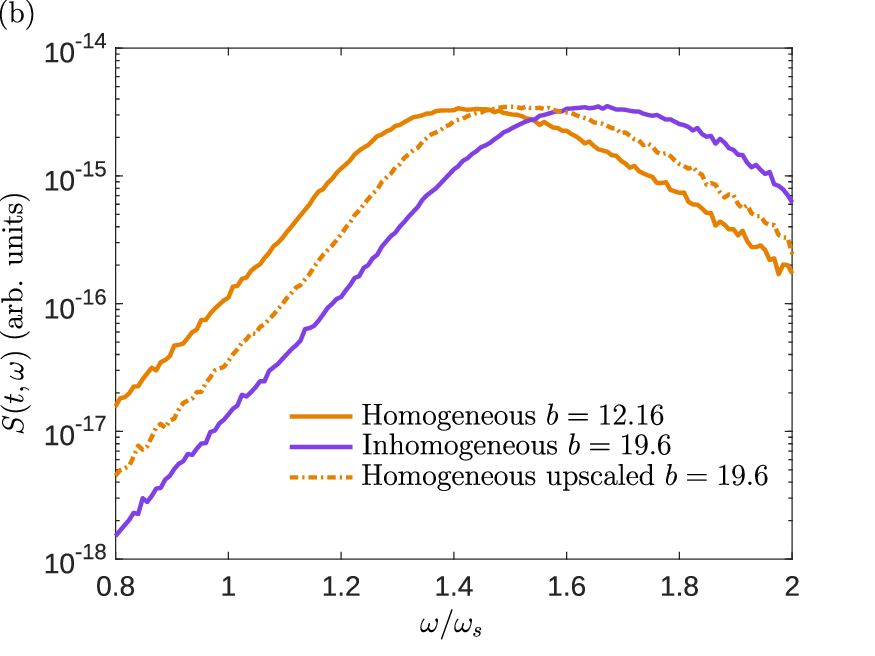}
      \includegraphics[width=0.95\linewidth]{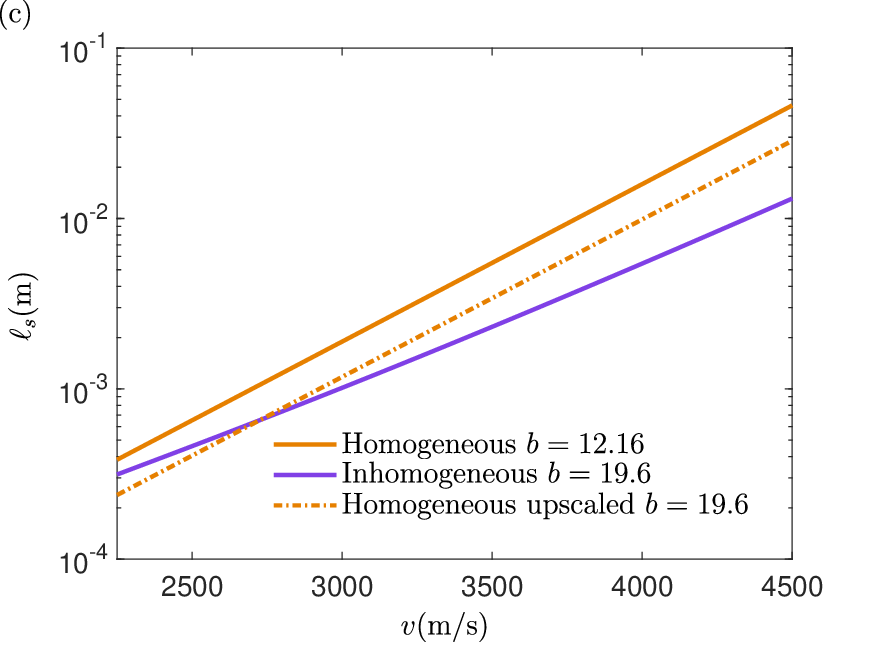}
      \caption{(a)~Comparison between the particle size distributions at the front of the ejecta (dotted line), 
      at the back (dashed-dotted line), and for a spatial average over the ejecta (solid line). 
      The parameters are given in Sec.~\ref{size_distributions_inhomogeneities}. 
      (b)~Comparison between ejecta layouts. Same parameters as in Fig.~\ref{holographie}\,(a), except for the 
      upscaled homogeneous spectrum obtained by increasing $M_s$ in order to get an optical thickness $b=19.6$. 
      (c)~Comparison between the scattering mean free paths for the three ejecta corresponding to the spectra in 
      Fig.~\ref{holographie}\,(b).}
      \label{holographie}
   \end{figure}

   Numerical simulations using this extended model are presented in Fig.~\ref{holographie}.
   In Fig.~\ref{holographie}\,(a), we plot the size distributions at the front of the ejecta, at the back, and for
   a spatial average over the ejecta. We see that the inhomogeneous distributions are quite narrow compared to 
   the distribution averaged over velocities, specifically at the front of the ejecta, and that they
   do not overlap much. This supports the idea that accounting for spatial inhomogeneity might be important. 
   Indeed, the spectra shown in Fig.~\ref{holographie}\,(b) exhibit substantial differences. First, the real optical thickness 
   in the inhomogeneous case is almost twice that obtained assuming a homogeneous distribution. This results from
   a rearrangement of matter in the ejecta that creates a dense and highly scattering region of small particles at the front, 
   which screens the back part of the ejecta.
   While this could have been only the effect of a higher optical thickness, the homogeneous upscaled configuration in
   Fig.~\ref{holographie}\,(b) shows the opposite. Even at equal optical thicknesses, the visibility of the fastest particle
   at the front remains higher in the inhomogeneous case. This can be understood with the examination of 
   Fig.~\ref{holographie}\,(c), where we plot the scattering mean-free path as a function of velocity. While shifting down
   the mean free paths using $M_s$ as a tuning parameter in order to reach the same optical thicknesses, the difference
   in slope of the $\ell_s$ curve keeps a smaller mean free path at the front of the ejecta in the inhomogeneous case. 
   In other words, the dense layer of small particles at the front of the ejecta generates a higher optical thickness. This
   effect is overlooked in models assuming homogeneous particle size distributions, introducing a potentially substantial
   bias in the interpretation of the spectrograms.
   
   \section{Conclusion}
   % ==================

   In summary, we have developed a model for PDV measurements in shock experiments, emphasizing the influence
   of multiple scattering of the probe light. Starting from first principles of wave scattering, we have established rigorously
   the relationship between the specific intensity and the measured signal in PDV experiments, as well as the RTE that 
   describes the evolution of the specific intensity upon scattering and absorption in the ejecta. The resulting RTE accounts 
   for inelastic scattering (Doppler shifts) and inhomogeneities in the particle density, size and velocity in the quasi-homogeneous
   regime. The derivation provides a rigorous basis to the RTE model for PDV experiments, going beyond the usual 
   phenomenological description, thus clarifying the underlying assumptions and limits of validity. We have also justified
   the use of a Monte-Carlo approach for numerical simulations, connecting the random-walk picture to the solution to
   the integral form of the RTE. Finally, we have used numerical simulations of realistic shock ejecta experiments to highlight 
   the role of multiple scattering, and of inhomogeneities in the particle density and size-velocity distribution induced by 
   the fragmentation process. 
   We have shown that linking inhomogeneous source terms to PDV spectra is possible, and should be applied to 
   other data sets than holographic measurements. In addition, since real ejecta have large optical thicknesses ($b>10$),
   a diffusion approximation of the RTE could be derived in order to simplify the analyses in practice. 
   These are potential lines to be followed in further investigations.

   \begin{acknowledgments}
      This work has received support under the program ``Investissements d’Avenir'' launched by the French Government. 
   \end{acknowledgments}

   \section*{Data Availability Statement}
   % ====================================
   The data that supports the findings of this study are available within the article.

   % Biblio
   \nocite{*}

\end{document}